%% file: nonflowFluct.tex
\pdfoutput=1
\documentclass[aps,prc,showpacs,twocolumn,superscriptaddress,preprintnumbers,amsmath,amssymb]{revtex4}
\usepackage{graphicx}
\usepackage{dcolumn}
\usepackage{bm}
\usepackage{color}

\definecolor{orange}{cmyk}{0.,0.353,1.,0.}    
\definecolor{dgreen}{cmyk}{1.,0.,1.,0.4}	  

\newcommand{\pt}{\ensuremath{p_T} }
\newcommand{\mean}[1]{\left\langle #1 \right\rangle}
\def \GeVc {\mbox{\ensuremath{\mathrm{GeV}/c} }}
\newcommand{\sqrtsNN}{\mbox{$\sqrt{\mathrm{s}_{_{\mathrm{NN}}}}$}}
\newcommand{\eps }{\varepsilon}
\def \vtep {\ensuremath{v_2\{\mathrm{EP}\} }}
\def \vtsep {\ensuremath{v_2\{\mathrm{subEP}\} }}
\def \vtpp {\ensuremath{v_{2,\mathrm{PP}} }}
\def \vtrp {\ensuremath{v_{2,\mathrm{RP}} }}
\def \vep {\ensuremath{v\{\mathrm{EP}\} }}
\def \vsep {\ensuremath{v\{\mathrm{subEP}\} }}
\def \vpp {\ensuremath{v_{\mathrm{PP}} }}
\def \vrp {\ensuremath{v_{\mathrm{RP}} }}

\def \npart {\ensuremath{N_{\mathrm{part}}}}
\def \nbin {\ensuremath{N_{\mathrm{bin}}}}
\def \res {{\mathcal{R}}}
\def \ch {{\mathcal{C}}}
\def \sigtot {\ensuremath{\sigma_{\mathrm{tot}}}}

\def \epspart {\ensuremath{\eps_{\mathrm{part}}}}

\begin{document}

\title{Effect of flow fluctuations and nonflow on elliptic flow methods}
\author{Jean-Yves Ollitrault}
\affiliation{CNRS, URA2306, Institut de physique th\'eorique de Saclay,
F-91191 Gif-sur-Yvette, France}
\author{Arthur M. Poskanzer}
\affiliation{Lawrence Berkeley National Laboratory, Berkeley,
California, 94720}
\author{Sergei A. Voloshin}
\affiliation{Wayne State University, Detroit, Michigan, 48201}

\date{\today}

\begin{abstract}
We discuss how the different estimates of elliptic flow are influenced by flow fluctuations and nonflow effects. It is explained why the event-plane method yields estimates between the two-particle correlation methods and the multiparticle correlation methods. It is argued that nonflow effects and fluctuations cannot be disentangled without other assumptions. However, we provide equations where, with reasonable assumptions about fluctuations and nonflow, all measured values of elliptic flow converge to a unique mean $\vtpp$ elliptic flow in the participant plane and, with a Gaussian assumption on eccentricity fluctuations,  can be converted to the mean $\vtrp$ in the reaction plane. Thus, the 20\% spread in observed elliptic flow measurements from different analysis methods is no longer mysterious.
\end{abstract}


\maketitle


\input{intro}

\input{methods}

\input{analytic}

\input{numeric.tex}

\input{data.tex}

\input{summary.tex}

\section{Acknowledgments}
We thank the authors of Ref.~\cite{Alver:2008zz} and Constantin Loizides for permission to use Fig.~\ref{fig:PHOBOS}. For discussions we thank Hiroshi Masui, Aihong Tang and Paul Sorensen. This work was supported in part by the HENP Divisions of the Office of Science of the U.S. Department of Energy under Contract Numbers DE-AC02-05CH11231 and DE-FG02-92ER40713.

\input{appendix}


\input{references}
\end{document}

%% file: intro.tex
\section{Introduction}
\label{sec:intro}

Elliptic flow has proved to be very valuable for understanding relativistic nuclear collisions~\cite{Voloshin:2008dg,Sorensen:2009}. However, different analysis methods give results which spread over a range of 20\%~\cite{Adams:2004bi}. A higher accuracy is now needed because when comparing to relativistic viscous hydrodynamic calculations, an uncertainty of 30\% in the elliptic flow parameter $v_2$ leads to an uncertainty of 100\% in the ratio of shear viscosity to entropy~\cite{Song:2008hj}. The experimental measurements need to converge to allow extraction of such important characteristics of the matter produced in relativistic nuclear collisions. The problem of nonflow correlations contributing to $v_2$ has been known for a long time~\cite{Borghini:2000cm}. More recently it has been recognized that fluctuations affect the measured $v_2$ values~\cite{{Miller:2003kd},Manly:2005zy}. It is now also recognized that some measurements are relative to the participant plane and some to the reaction plane~\cite{Voloshin:2007pc}.
The \textit{reaction plane} is spanned by the vector of the impact parameter and the beam direction. The participants are those constituents which partake in the primary interaction. The minor axis of the participant zone and the beam direction define the \textit{participant plane}. The \textit{event plane} contains the flow vector $Q$ constructed from the transverse momenta of the detected particles.

The purpose of this paper is to propose a method using reasonable assumptions to obtain a well-defined measure of elliptic flow to compare with theoretical calculations. Section~\ref{sec:methods} describes the flow analysis methods we will be discussing. The analytic equations are derived in Secs.~\ref{sec:analytic} and \ref{sec:nonflow}, and summarized in Sec.~\ref{sec:disc}. Tests of the equations by numerical integrations and simulations are in Sec~\ref{sec:numeric}. Then the analytic equations are applied to published STAR data in Sec.~\ref{sec:data}. Section~\ref{sec:summary} is a summary.

For simplicity we will write $v\{\ \}$ instead of $v_n\{\ \}$ and $\cos(...)$ instead of $\cos[n(...)]$, where $n$ is the harmonic number of the anisotropic flow. The final equations are independent of $n$.

%% file: methods.tex
\section{Flow methods}
\label{sec:methods}

The two-particle cumulant method $v\{2\}$ correlates each particle with every other particle, and is defined as~\cite{Wang:1991qh}
\begin{equation}
  v\{2\}\equiv\sqrt{\langle\cos(\phi_1-\phi_2)\rangle} \label{defv22} \ ,
\end{equation}
where $\mean{\ }$ indicates an average over all particles in all events.
The four-particle cumulant method $v\{4\}$ is defined as~\cite{Borghini:2001vi}
\begin{equation}
  v\{4\}\equiv\left(2\langle\cos(\phi_1-\phi_2)\rangle^2-
  \langle\cos(\phi_1+\phi_2-\phi_3-\phi_4)\rangle\right)^{1/4}.
\label{defv24}
\end{equation}
The Lee-Yang Zeros method~\cite{Borghini:2004ke} $v\{{\rm LYZ}\}$ is also a multiparticle correlation.

The event-plane method $\vep$ correlates each particle with the event plane of the other particles. The event-plane azimuth $\Psi_R$ is defined as the azimuthal angle of the flow vector:
\begin{eqnarray}
  {\bf q}\cos\Psi_R =\frac{\bf Q}{\sqrt{N}}\cos\Psi_R&=&\frac{1}{\sqrt{N}}\sum_{j=1}^N \cos\phi_j \cr
  {\bf q}\sin\Psi_R=\frac{\bf Q}{\sqrt{N}}\sin\Psi_R &=&\frac{1}{\sqrt{N}}\sum_{j=1}^N \sin\phi_j,
\label{defq}
\end{eqnarray}
where $|q| \ge 0$ and the sum runs over particles defining the event plane. Since $Q$, the magnitude of the standard flow vector, is proportional to $\sqrt{N}$ in the absence of correlations, it is convenient to use ${\bf q} \equiv {\bf Q}/\sqrt{N}$.
The event-plane estimate of anisotropic flow is defined as 
\begin{equation}
  v\{{\rm EP}\}\equiv \frac{\langle\cos(\phi-\Psi_R)\rangle}{R},
\label{defvep}
\end{equation}
where the particle of interest is always subtracted from ${\bf q}$ before calculating $\Psi_R$ to avoid autocorrelations. $R$ is the event plane {\it resolution\/} correction which is determined from the correlation between the event plane vectors of two independent ``subevents'' $A$ and $B$. 
In the original method~\cite{Danielewicz:1985hn,Poskanzer:1998yz},
subevent $A$ is defined by choosing randomly $N/2$ particles out of
the $N$ particles of the event plane and subevent $B$ is made of the 
remaining $N/2$ particles. Other methods of choosing the subevents are now also used, such as according to pseudorapidity or charge, or combinations of these. For sake of generality, we denote by $N_s$ the number of particles in a subevent. The azimuths $\Psi_A$ and $\Psi_B$ are defined by equations similar to
Eq.~(\ref{defq}), where $N$ is replaced with $N_s$. 

In the special case where the event plane comes from only one subevent, $N=N_s$, the resolution correction is the subevent resolution: 
\begin{equation}
  R=\sqrt{\langle\cos(\Psi_A-\Psi_B)\rangle} \ . 
\label{subresolution}
\end{equation}
The corresponding estimate of anisotropic flow will be denoted by
\vsep, or, more particularly, $v\{{\rm etaSub}\}$ or $v\{{\rm ranSub}\}$, depending on how the events were divided. 

In the more general case when the event plane comes from the full event, one first estimates the resolution parameter $\chi_s$ of 
the subevents by solving numerically the equation
\begin{equation}
  {\cal R}(\chi_s)= \sqrt{\langle\cos(\Psi_A-\Psi_B)\rangle} \ ,
\label{defchis}
\end{equation}
where the function ${\cal R}$ is defined by~\cite{Poskanzer:1998yz,Ollitrault:1997di,Ollitrault:1997vz} 
\begin{equation}
  {\cal R}(\chi)=\frac{\sqrt{\pi}}{2}e^{-\chi^2/2}\chi\left(
  I_0\left(\frac{\chi^2}{2}\right)+I_1\left(\frac{\chi^2}{2}\right)
  \right),
\label{resolution}
\end{equation}
where $I_0$ and $I_1$ are modified Bessel functions. 
Generally, the resolution parameter is related to the flow through
\begin{equation}
  \chi_s=v\sqrt{N_s} \ .
\label{chiversusn}
\end{equation}
One then estimates the resolution parameter $\chi$ of the full event
as $\chi\equiv \chi_s\sqrt{N/N_s}$. The resolution correction for the full event $R$ is defined by 
\begin{equation}
  R\equiv {\cal R}(\chi)={\cal R}(\chi_s\sqrt{N/N_s}) \ .
\label{defresolution}
\end{equation} 
If the event plane coincides with one subevent
$\chi=\chi_s$, and Eqs.~(\ref{defchis}) and (\ref{defresolution})
reduce to Eq.~(\ref{subresolution}). 

For a review of anisotropic flow see Ref.~\cite{Voloshin:2008dg}.

%% file: analytic.tex
\section{Fluctuations}
\label{sec:analytic}

Elliptic flow is driven by the initial eccentricity of the overlap
almond~\cite{Ollitrault:1992bk}. This eccentricity fluctuates from one event to the other. There are several sources of fluctuations: fluctuations of
impact parameter within the sample of events~\cite{Adler:2002pu} and,
more importantly, fluctuations of the positions of participant 
nucleons~\cite{Miller:2003kd,Manly:2005zy,Alver:2006wh}. It is
fluctuations which make $\mean{v}$ in the participant plane larger
than in the reaction plane. The magnitude of flow fluctuations is
characterized by $\sigma_v$, defined by 
\begin{equation}
\sigma_v^2\equiv\mean{v^2}-\mean{v}^2,
\label{defsigmav}
\end{equation} 
where $v$ is the flow in the participant plane $\vpp$ in the case of fluctuations in the participant plane.
Flow methods involve various functions of $v$, which are also affected
by fluctuations. The average value of $f(v)$ is obtained by expanding
around $\mean{v}$ to leading order in $\sigma_v^2$:
\begin{equation}
\mean{f(v)}= f(\mean{v})+\frac{\sigma_v^2}{2}f''(\mean{v}).
\label{averagef}
\end{equation}
This result will be useful below.

We now derive the effect of fluctuations on the various flow
estimates, to order $\sigma_v^2$. 
Using the definitions of $v\{2\}$ and $v\{4\}$ from Eqs.~(\ref{defv22})
and (\ref{defv24}), we have
\begin{equation}
v\{2\}^2= \mean{v^2}=\mean{v} ^2 +\sigma_v^2
\label{fluctv2}
\end{equation}
and
\begin{eqnarray}
v\{4\}^2&=& \left(2\mean{v^2}^2 -\mean{v^4}\right)^{1/2}\cr
&\approx &\mean{v}^2-\sigma_v^2 \ .
\label{fluctv4}
\end{eqnarray}
Fluctuations increase $v\{2\}$ and decrease $v\{4\}$ compared to $\vpp$. 
In the case of Gaussian eccentricity fluctuations, $v\{4\}$
measures the correlation to the true reaction plane $\vrp$
\cite{Voloshin:2007pc,Bhalerao:2006tp}. However, it has been shown 
that eccentricity fluctuations are not quite Gaussian, especially for peripheral collisions~\cite{Voloshin:2007pc, Alver:2008zz}. 

The contribution of fluctuations to the various $v\{\ \}$ results can be parametrized by $\alpha$~\cite{Alver:2008zz}:
\begin{equation}
v\{\ \} = \mean{v^\alpha}^{1/\alpha}.
\label{alphaphobos}
\end{equation}
Equation~(\ref{averagef}) with $f(v)=v^\alpha$ gives 
\begin{equation}
\mean{v^\alpha}=\mean{v}^\alpha\left(1+\frac{\sigma_v^2}{\mean{v}^2}\frac{\alpha(\alpha-1)}{2}\right). 
\end{equation}
Raising to the power $2/\alpha$ and expanding to leading order in
$\sigma_v^2$, one gets
\begin{equation}
v\{\ \}^2 = \mean{v}^2 + (\alpha-1)\sigma_v^2 \ .
\label{defalpha} 
\end{equation}
Note that $v\{4\}$ from Eq.~(\ref{fluctv4}) corresponds to
the limiting case $\alpha=0$ and $v\{2\}$ from Eq.~(\ref{fluctv2}) corresponds to the case $\alpha=2$. The event plane methods have intermediate $\alpha$.

We now derive the value of $\alpha$ for \vsep, defined by  
Eqs.~(\ref{defvep}) and (\ref{subresolution}). 
The subevent resolution depends on the flow $v$, which fluctuates:
\begin{equation}
\label{subAverage}
\vsep^2=\frac{\mean{v {\cal R}(v)}^2}{\mean{ {\cal R}^2(v)}} \ .
\end{equation}
The averages in the numerator and in the denominator can be evaluated
by using Eq.~(\ref{averagef}). Expanding to leading order in $\sigma_v^2$, 
one obtains 
\begin{equation}
\vsep^2=\mean{v}^2+
\frac{\mean{v}{\cal R}'}{\cal R}\left[2-\frac{\mean{v}{\cal R}'}{\cal
    R} \right]\sigma_v^2 \ , 
\end{equation}
where ${\cal R}'$ is the derivative of ${\cal R}$ with respect to
$v$. 
Comparing to Eq.~(\ref{defalpha}), one obtains the following
  expression of $\alpha$, which is independent of $\sigma_v$:
\begin{equation}
\alpha = 1+ \frac{\mean{v}{\cal R}'}{\cal R}\left[2-\frac{\mean{v}{\cal R}'}{\cal R} \right].
\end{equation}
Inserting Eq.~(\ref{defresolution}) for ${\cal R}(\chi)$
and using the fact that $\chi$ is
proportional to $v$, one obtains after some algebra
\begin{equation}
  \alpha = 2 - \frac{4 i_1^2}{(i_0+i_1)^2} \ ,
\label{alphasubep}
\end{equation}
where $i_{0,1}$ is a shorthand notation for $I_{0,1}(\chi_s^2/2)$.

\begin{figure}[hbt]
\begin{center}
\centerline{\includegraphics[width=.48\textwidth]{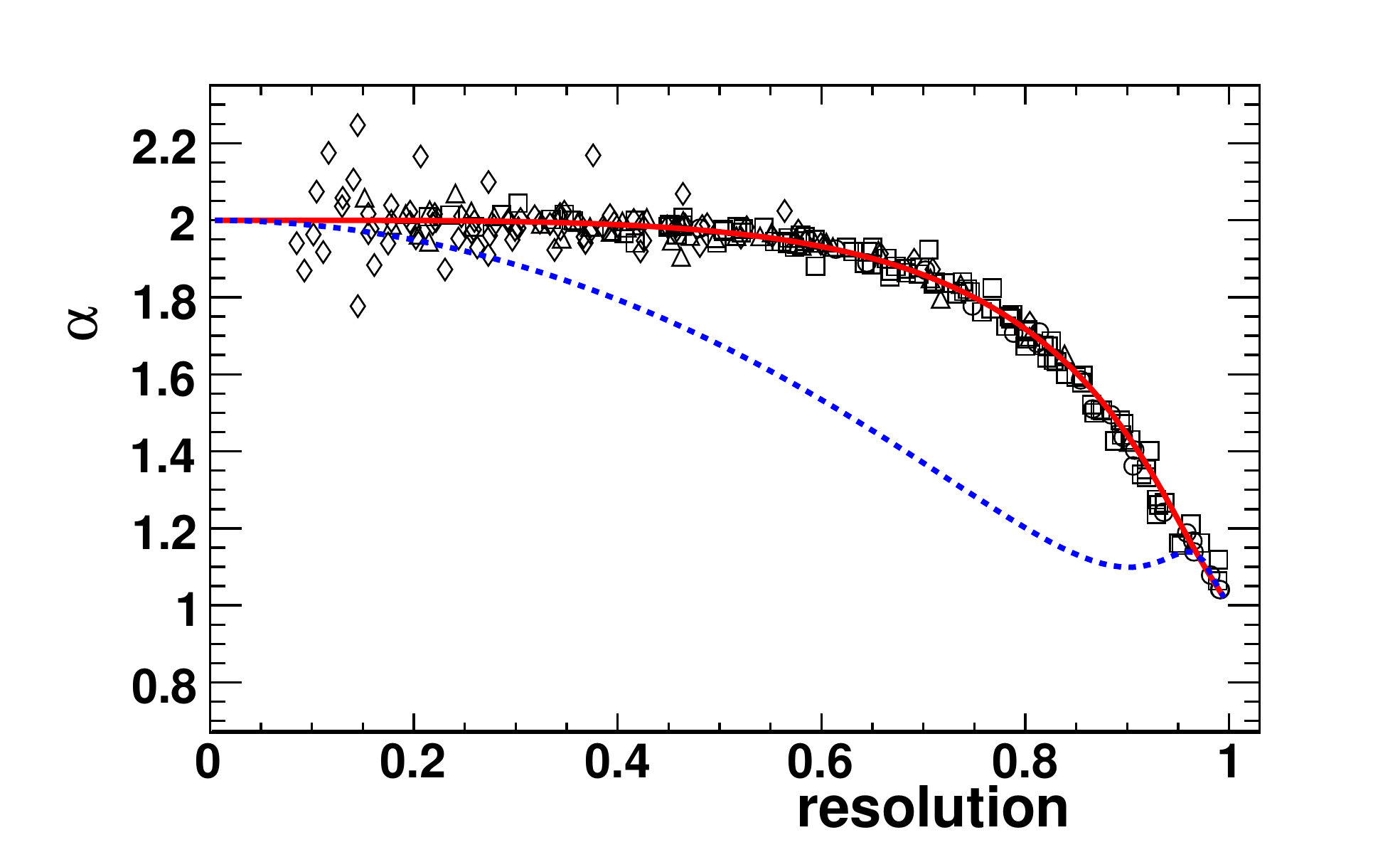}}
  \caption{(Color online) Alpha, defined in Eq.~(\ref{alphaphobos}), vs the subevent plane resolution. The simulations of $\vtsep$ are from Ref.~\cite{Alver:2008zz}. The solid line is Eq.~(\ref{alphasubep}). The dashed line is Eq.~(\ref{alphaep}) plotted vs the full event plane resolution. }
  \label{fig:PHOBOS}
\end{center}
\end{figure}

As an example of the application of Eq.~\ref{alphasubep} we re-plot Ref.~\cite{Alver:2008zz} Fig.~5 as Fig.~\ref{fig:PHOBOS} here. Alpha is defined by Eq.~(\ref{alphaphobos}) and the resolution is the subevent plane resolution. Simulations including event-by-event fluctuations were done and analyzed with the subevent method, and using Eq.~(\ref{alphaphobos}) alpha was extracted. Our Eq.~(\ref{alphasubep}) has been added to the figure without any adjustable parameters. The extraordinary fit means that fluctuations quantitatively explain the figure.

Inserting Eq.~(\ref{alphasubep}) into Eq.~(\ref{defalpha}), one gets
\begin{equation}
\vsep^2
=\mean{v}^2+\left(1-\frac{4
   i_1^2}{(i_0+i_1)^2}\right)\sigma_v^2 \ .
\label{flvsubep}
\end{equation}
The case where the event plane consists of two subevents is studied in
Appendix~\ref{sec:appendixA}.  
Equation~(\ref{flvsubep}) is replaced by the more general expression 
\begin{eqnarray}
  \lefteqn{\vep^2 = \mean{v}^2} \cr
  &+& \left(1-\frac{I_0-I_1}{I_0+I_1}\left(2\chi^2-2\chi_s^2+\frac{4
   i_1^2}{i_0^2-i_1^2}\right)\right)\sigma_v^2 \ ,
\label{flvep}
\end{eqnarray}
where again $i_{0,1}$ is a shorthand notation for
$I_{0,1}(\chi_s^2/2)$ for subevents, 
and $I_{0,1}$ is a shorthand notation for $I_{0,1}(\chi^2/2)$
for the whole event.
The corresponding expression for
$\alpha$ is obtained again by comparing to Eq.~(\ref{defalpha}): 
\begin{equation}
\alpha=2
-\frac{I_0-I_1}{I_0+I_1}\left(2\chi^2-2\chi_s^2+\frac{4
   i_1^2}{i_0^2-i_1^2}\right). 
\label{alphaep}
\end{equation}
When the event plane consists of one subevent only, $\chi=\chi_s$,
$I_{0,1}=i_{0,1}$, and Eqs.~(\ref{flvep}) and (\ref{alphaep}) reduce
to Eqs.~(\ref{flvsubep}) and (\ref{alphasubep}), respectively. Equation~(\ref{alphaep}) is also plotted in Fig.~\ref{fig:PHOBOS}. It is lower than Eq.~(\ref{alphasubep}) and explains why \vtep\ is generally lower than \vtsep.

\section{Nonflow effects}
\label{sec:nonflow}

In this section we discuss nonflow effects while neglecting
fluctuations. $v$ can therefore be identified with $\mean{v}$. 
The two-particle azimuthal correlation gets contributions from flow
and from other, ``nonflow'' effects: 
\begin{equation}
  \mean{\cos(\phi_1-\phi_2)} \equiv \mean{v} ^2+\delta \ ,
\label{delDef}
\end{equation}
where $\delta$ is the nonflow part. 
One expects that $\delta$ varies with centrality like $1/N$,
where $N$ is some measure of the multiplicity~\cite{Poskanzer:1998yz,Borghini:2000cm}. 

Using Eqs.~(\ref{defv22}) and (\ref{delDef}), one obtains, to leading
order in $\delta$,
\begin{equation}
v\{2\}^2=\mean{v}^2+\delta
\label{nonflowv2}
\end{equation}
On the other hand, $v\{4\}$ is insensitive to nonflow effects, thus
\begin{equation}
v\{4\}=\mean{v} \ .
\label{nonflowv4}
\end{equation}

We now derive the expression of $\vep$ to leading order in $\delta$. 
In the same way as fluctuations, nonflow effects contribute to both
the numerator and denominator of Eq.~(\ref{defvep}). 
These contributions are evaluated in detail in 
Appendix~\ref{sec:appendixB}. The nonflow correlation between the
particle and the event plane (numerator) is derived by shifting the
flow vector by an amount proportional to $\delta$ and to the unit
vector of the particle. The nonflow correlation between subevents is
taken into account in the probability distribution of ${\bf q_A}$,
${\bf q_B}$ by a correlation term, whose form is dictated by the
central limit theorem. One must also take into account the fact that
nonflow correlations modify the width of fluctuations of the flow
vector around the reaction plane, which are responsible for the
resolution correction in Eq.~(\ref{defvep}). One obtains
\begin{eqnarray}
\lefteqn{\vep^2 = \mean{v}^2} \cr
&+& \left(1-\frac{I_0-I_1}{I_0+I_1}\left(\chi^2-\chi_s^2+\frac{2
 i_1^2}{(i_0^2-i_1^2)}\right) 
\right)\delta \ .
\label{nonflowep}
\end{eqnarray}
If the event plane consists of only one subevent, $I_{0,1}=i_{0,1}$,
$\chi=\chi_s$, and this simplifies to
\begin{equation}
\vsep^2=
\mean{v}^2
+\left(1-\frac{2i_i^2}{(i_0+i_1)^2}\right)\delta \ . 
\label{nonflowsubep}  
\end{equation}
If the resolution is low, $i_1\ll i_0$ and $\vsep$ coincides with
$v\{2\}$, Eq.~(\ref{nonflowv2}). If the resolution is large,
$i_1\simeq i_0$ and $\vsep$ lies half-way between $v\{2\}$ and $v\{4\}$.

\section{Summary of equations}
\label{sec:disc}

We assume that to leading order in $\sigma_v^2$ and $\delta$, the contributions of nonflow and fluctuations are additive. Equations~(\ref{fluctv2}) and (\ref{nonflowv2})
yield:
\begin{equation}
 v\{2\}^2=
 \langle v\rangle^2 +\delta+\sigma_v^2 \ .
\label{sumv2}
\end{equation}
Similarly, Eqs.~(\ref{fluctv4}) and (\ref{nonflowv4}) yield 
\begin{equation}
 v\{4\}^2=
 \langle v\rangle^2 -\sigma_v^2 \ .
\label{sumv4}
\end{equation}
Although this equation was derived for $v\{4\}$ it should apply to all multiparticle values. As for the event-plane method, Eqs.~(\ref{flvep}) and (\ref{nonflowep})
give
\begin{widetext}
\begin{equation}
 v\{{\rm EP}\}^2=\langle
 v\rangle^2+
 \left(1-\frac{(I_0-I_1)}{(I_0+I_1)}\left(\chi^2-\chi_s^2+\frac{2 i_1^2}{
 (i_0^2-i_1^2)}\right)\right)\delta
 +\left(1-\frac{2(I_0-I_1)}{I_0+I_1}\left(\chi^2-\chi_s^2+\frac{2
 i_1^2}{i_0^2-i_1^2}\right)\right)\sigma_v^2 \ .
\label{sumvep}
\end{equation}
Finally, Eqs.~(\ref{flvsubep}) and (\ref{nonflowsubep}) yield
\begin{equation}
 v\{{\rm subEP}\}^2= \langle v\rangle^2 +
 \left(1-\frac{2 i_1^2}{
 (i_0+i_1)^2}\right)\delta
 +\left(1-\frac{4\,i_1^2}{(i_0+i_1)^2} \right)
 \sigma_v^2 \ . 
\label{sumvsub}
\end{equation}
This can be derived from Eq.~(\ref{sumvep}) by setting $\chi = \chi_s$.

The difference between estimates always scales like 
$\delta+2\sigma_v^2$:
\begin{eqnarray}
  v\{2\}^2-v\{4\}^2&=& \delta+2\sigma_v^2\cr 
  v\{2\}^2-v\{{\rm EP}\}^2&=&
\frac{(I_0-I_1)}{(I_0+I_1)}\left(\chi^2-\chi_s^2+\frac{2i_1^2}{(i_0^2-i_1^2)}\right) \left(\delta+2\sigma_v^2\right)\cr 
\cr
  v\{2\}^2-v\{{\rm subEP}\}^2&=& \frac{2 i_1^2}{(i_0+i_1)^2}
  \left(\delta+2\sigma_v^2\right) \ .
\label{differences}
\end{eqnarray}
\end{widetext}
Thus we have defined $\sigma_{\rm tot}^2 \equiv \delta + 2 \sigma_{v}^2$. This shows explicitly that fluctuations and nonflow effects cannot be disentangled with only these measurements.

%% file: numeric.tex
\section{Numeric integrations and simulations}
\label{sec:numeric}

To avoid the leading order expansions in Sec.~\ref{sec:analytic}, one can test the accuracy of the analytic equations by performing numeric integrations or analyzing simulations to solve for the various $v\{\ \}$ quantities from $\mean{v}$ and $\sigma_v$ in the participant plane. If one assumes a Gaussian distribution there is a tail to negative values of $v$ for large fluctuations. Since participant eccentricity never goes negative, one can avoid this problem by using a Bessel-Gaussian distribution. Also, assuming a two-dimensional Gaussian in the reaction plane makes the distribution along the participant plane axis have the form of a Bessel-Gaussian~\cite{Voloshin:2007pc}:
\begin{equation}
  \frac{dn}{dv} = \frac{v}{\sigma_{0}^2} 
  I_0 \left( \frac{v \ v_0 }{\sigma_{0}^2} \right) 
  \exp \left( -\frac{v^2 + v_0^2}{2\sigma_{0}^2} \right) \ ,
\label{BG}
\end{equation}
where $v_0$ and $\sigma_{0}$ are parameters of the distribution which
are adjusted so that the first and second moments equal $\mean{v}$ and
$\sigma_v$. The equations for these moments are in
Ref.~\cite{Voloshin:2007pc}. The relative magnitude of fluctuations is maximum for $v_0=0$, corresponding to zero impact parameter central collisions:
  $\sigma_v/\mean{v}=\sqrt{(4/\pi)-1}=52.2\%$~\cite{Broniowski:2007ft}.

\subsection{Numeric integrations}
\label{sec:num}
\begin{figure}[hbt]
\centerline{\includegraphics[width=.48\textwidth]{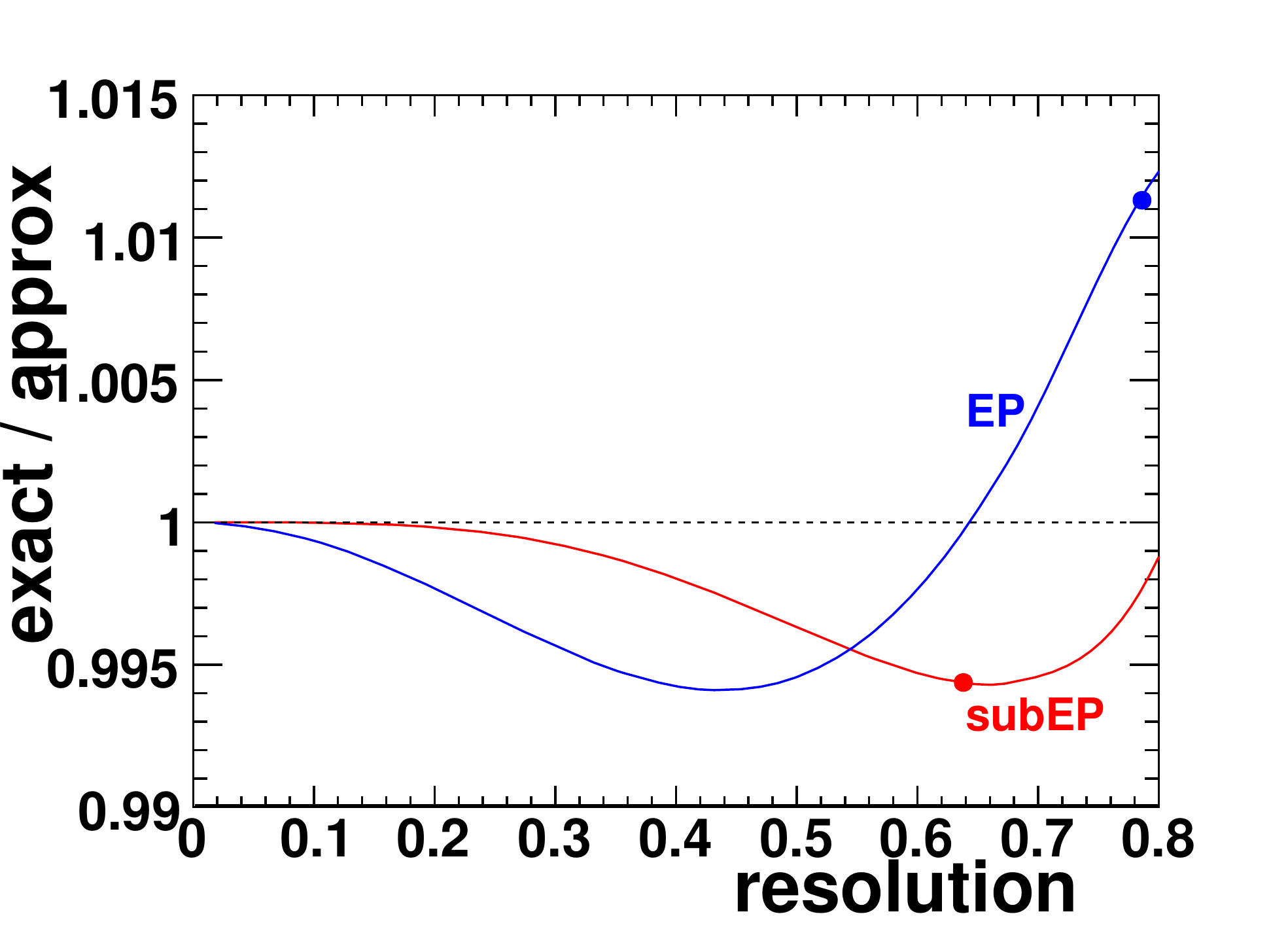}}
\caption{(Color online) Ratio of the exact values of $\vsep$ and $\vep$,
  defined by Eqs.~(\ref{subs}) and (\ref{EP}), to the approximate expressions (\ref{flvsubep}) and (\ref{flvep}), for central collisions, vs the reaction plane resolution. The $x$-axis is the full event resolution for \vep\ and the subevent resolution for \vsep. The points correspond to the points in Fig.~\ref{fig:anaNumSig}. However, for central collisions the resolution is normally lower than this.}
\label{fig:b=0}
\end{figure}

For the subevent plane and full event plane flow values we evaluate from Eq.~(\ref{subAverage})
\begin{equation}
  v\{{\rm subEP}\} = \frac{\langle v \ \res(v \ \sqrt{N/2}) \rangle} {\sqrt{\langle [ \res (v \  \sqrt{N/2})]^2 \rangle}}
\label{subs}
\end{equation}
\begin{equation}
 v\{{\rm EP}\} = \frac{ \langle v \ \res(v \ \sqrt{N}) \rangle } { \res \!\! \left[ \ch \!\! \left( \sqrt{\langle [ \res (v \ \sqrt{N/2})]^2 \rangle} \right) \sqrt{2} \right] } ,
\label{EP}
\end{equation}
where $\res$ and $\ch$ are functions which calculate the event plane
resolution and resolution parameter $\chi$, respectively. $\res$ is
given by Eq.~(\ref{resolution}) and $\chi$ is solved from that equation
by iteration. The averages are taken by integrating over the
normalized Bessel-Gaussians from 0 to $v_0 +4\ \sigma_{0}$.
For central collisions, $v_0=0$ and integrations can be done
analytically, as shown in Appendix~\ref{sec:appendixA2}. 
For this zero impact parameter case, Fig.~\ref{fig:b=0} displays the ratio of the exact values of $\vsep$ and  $\vep$ to the approximate expressions derived in  Sec.~\ref{sec:analytic}. This figure shows that, for realistic values of the resolution, the formulas in Sec.~\ref{sec:analytic} for maximum fluctuations are valid within 1\%.   

For the nonflow dependence, adding $\delta/2$ to $\sigma_v^2$ would only take into account the broadening of the distribution and not the direct nonflow correlations. Thus the effect of nonflow was tested only by simulations.

\subsection{Simulations}
\label{sec:sim}

The simulation results for fluctuations were obtained by generating 8 million events of fixed multiplicity $=400$ and elliptic flow values uniformly distributed in the range of 0 to 0.2. The angle of each track was selected randomly according to the azimuthal distribution defined by the elliptic flow of that event. After all tracks were generated they were divided into two
equal subevents and the corresponding flow vectors generated. The event
plane resolution, the observed flow, and the final flow values, were
calculated by applying a weight to each event according to a Bessel-Gaussian distribution with parameters which produced $\mean{v}=0.06$ and a corresponding $\sigma_v$ for plotting versus $\sigma_v/\mean{v}$.

The nonflow effects were simulated by generating similar events of fixed multiplicity $=200$ without flow and different numbers of pairs of particles with exactly the same azimuthal angle. If $f$ is the fraction of all particles generated as pairs with the same azimuth, $\delta = f/N$, where $N$ is the full multiplicity.

\subsection{Tests of the equations}
\label{sec:tests}
\begin{figure*}[hbt]
\begin{minipage}[t]{0.48\textwidth}
\includegraphics[width=.98\textwidth]{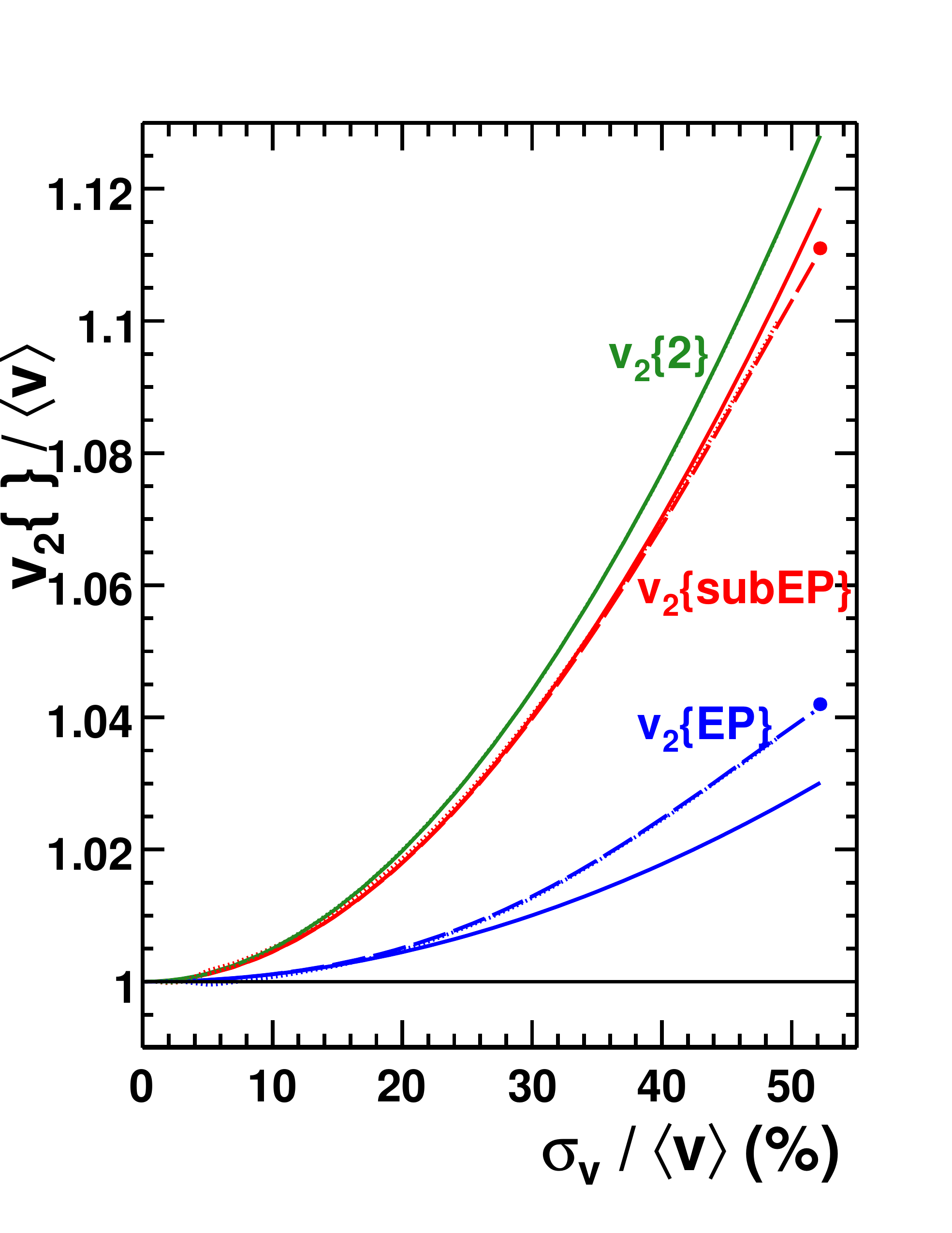}
  \caption{(Color online) Various $v\{\ \}$ values as a function of the the magnitude of the fluctuations calculated for $\mean{v}= 0.06$ with a full event multiplicity $= 400$ and $\delta = 0$. The solid curves are the analytic formulas. The points at $\sigma_v / \mean{v} = 52.2 \%$ are exact analytic calculations for zero impact parameter collisions. The dotted curves are from simulations. The dashed curves are the numerical integrations. The dotted curve for $v\{2\}$ is just under the solid curve. The dashed and dotted curves coincide for \vtsep \ (slightly lower than the solid line) and \vtep \ (above the solid line). Central collisions have large values of $\sigma_v / \mean{v}$.}
  \label{fig:anaNumSig}
\end{minipage}
\hspace{\fill}
\begin{minipage}[t]{0.48\textwidth}
\includegraphics[width=.98\textwidth]{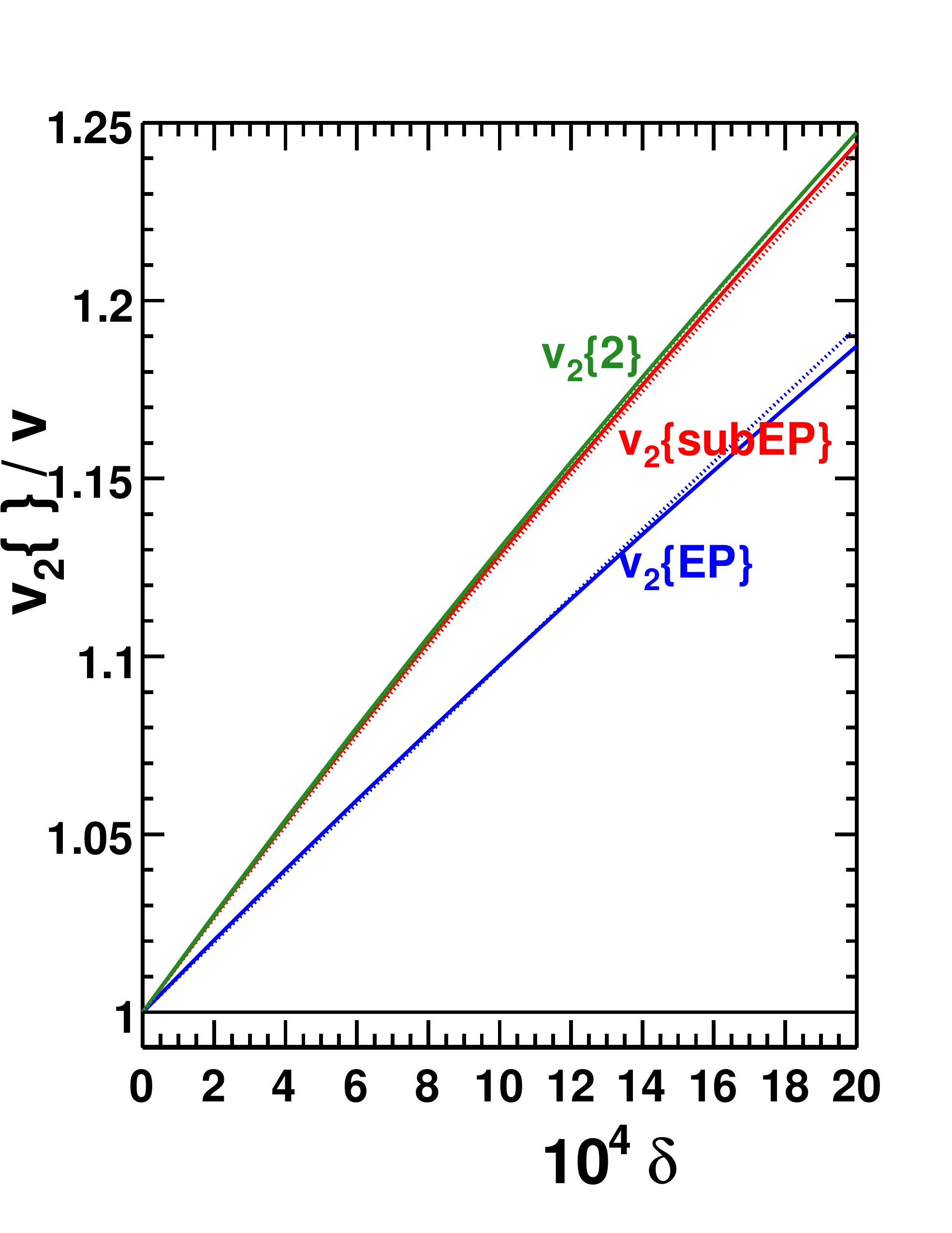}
  \caption{(Color online) Various $v\{\ \}$ values as a function of the the magnitude of nonflow calculated for $\mean{v}= 0.06$ with a full event multiplicity $= 200$ and $\sigma_v = 0$. The solid curves are the analytic formulas. The dotted curves are from simulations. The dotted curve for $v_2\{2\}$ coincides with the solid curve. For \vtsep \ the dotted curve is slightly below the solid curve, and for \vtep \ it is just above the solid curve. Peripheral collisions have large values of $\delta$.}
  \label{fig:anaNumDel}
\end{minipage}
\end{figure*}

For fixed $\mean{v}$ in the participant plane, the corrections for the
analytic method from Sec.~\ref{sec:disc}, the numeric method from
Sec.~\ref{sec:num}, and the simulation method from
Sec.~\ref{sec:sim} are given in Figs.~\ref{fig:anaNumSig} and
\ref{fig:anaNumDel}. The numerical and simulation methods agree
exactly. The values of $\sigma_v/\mean{v}$ that will be considered in
Sec.~\ref{sec:data} are shown in Table~\ref{tab:params} and go up
to about 50\% for the most central collisions. By using the exact equations in Appendix~\ref{sec:appendixA2}, points were plotted for these most central collisions in Fig.~\ref{fig:anaNumSig} at $\vsep / \mean{v} = 52.2\%$. They agree exactly with the numeric and simulation methods, validating those methods. Thus one can see that the approximations of the analytic equations for the fluctuation dependence are less than about 0.5\% for $\vsep / \mean{v}$ and 1.0\% for $\vep / \mean{v}$.
Since the $\delta$ values go up to about $20 \times 10^{-4}$ the approximations of the analytic equations for the nonflow dependence in Fig.~\ref{fig:anaNumDel} for $\vep / v$ and $\vsep / v$ are very small.

%% file: data.tex
\section{Application to data}
\label{sec:data}
So far the equations have used generic fluctuation and nonflow parameters. To apply the equations to data we now assume that the fluctuations arise from participant eccentricity fluctuations and that the nonflow is related to the elliptic flow in $p+p$ collisions scaled by the number of participants. Thus to apply the analytic equations in Sec.~\ref{sec:disc} to extract $\mean{v}$ in the participant plane from experimental data, we have assumed that the fluctuations in $v$ have the same fractional width as the fluctuations of the participant eccentricity:
\begin{equation}
  \sigma_{v} = \frac{\sigma_{\varepsilon}}{\langle \varepsilon \rangle} \langle v \rangle \ .
\label{sigma_v}
\end{equation}
Using this equation, Eq.~({\ref{sumv2}}) for $v\{2\}$ can be solved as
\begin{equation}
  \mean{v} = \sqrt{(v\{2\}^2 - \delta) / (1 + (\sigma_{\eps}/\mean{\eps})^2)}
\label{<v22>}
\end{equation}
and Eq.~({\ref{sumv4}}) for $v\{4\}$ as
\begin{equation}
  \mean{v} = v\{4\} / (1 + \sigma_{\eps}/\mean{\eps}) \ .
\label{<v24>}
\end{equation}
Because $\mean{v}$ appears in Eq.~(\ref{chiversusn}) for $\chi$, Eqs.~(\ref{sumvep}) and (\ref{sumvsub}) have to be solved by iteration.

\subsection{Glauber fluctuations}
\begin{figure}[hbt]
\begin{center}
\centerline{\includegraphics[width=.48\textwidth]{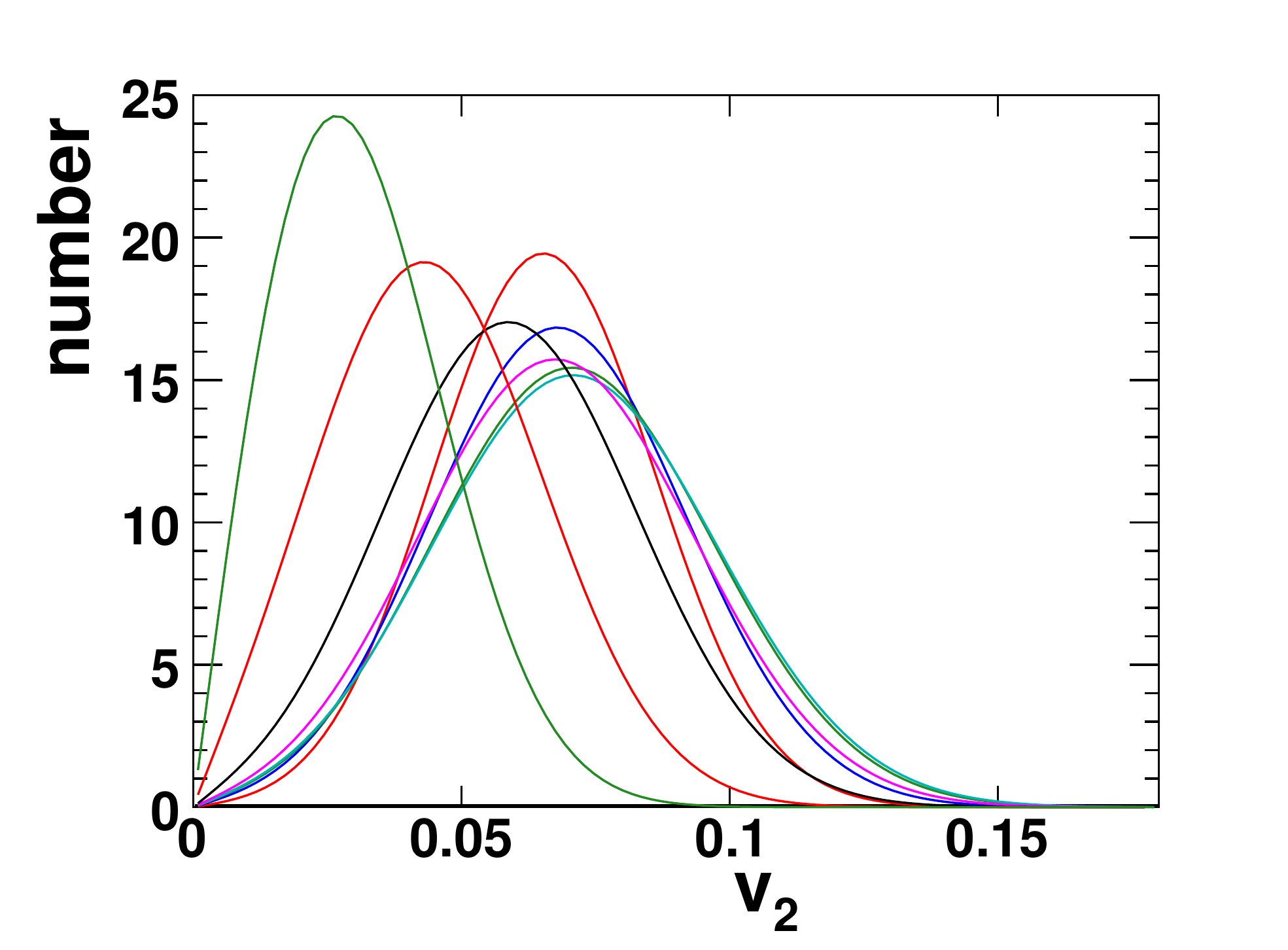}}
  \caption{(Color online) Bessel-Gaussian fluctuation distributions of $v$ assuming the same fractional width as $\epspart$ from Monte-Carlo Glauber calculations~\cite{Hiroshi}. The different curves correspond to the first eight centrality bins of STAR for $\sqrtsNN = 200$ GeV Au+Au. See Table~\ref{tab:params}. }
  \label{fig:BG}
\end{center}
\end{figure}
A nucleon Monte-Carlo Glauber calculation was used to calculate the fractional standard deviation of \epspart~\cite{Hiroshi}. The $\mean{v}$ values were calculated from $v\{2\}$ by using Eq.~(\ref{<v22>}) with $\delta=0$. Assuming Bessel-Gaussians, the resulting distributions are shown in Fig.~\ref{fig:BG}.

For the nonflow contribution we have taken the value from proton-proton collisions and scaled it down by the number of participants. The value of $\delta_{pp}$ was obtained by integrating the minimum bias $p+p$ curves of Ref.~\cite{Adams:2004wz}, Fig.~1, and it was found that $\delta_{pp} = 0.0145$~\cite{Tang}. Thus for nonflow as a function of centrality we assume 
\begin{equation}
  \delta = \delta_{pp} \ 2 / \npart \ ,
\label{delta}
\end{equation}
knowing that in a $p+p$ collision there are two participants. One could also scale with $1/\rm{multiplicity}$. Doing that we get as good results as shown below, but because multiplicity depends on acceptance, an extra parameter is needed.

\begin{figure*}[hbt]
\begin{minipage}[t]{0.48\textwidth}
 \includegraphics[width=0.98\textwidth]{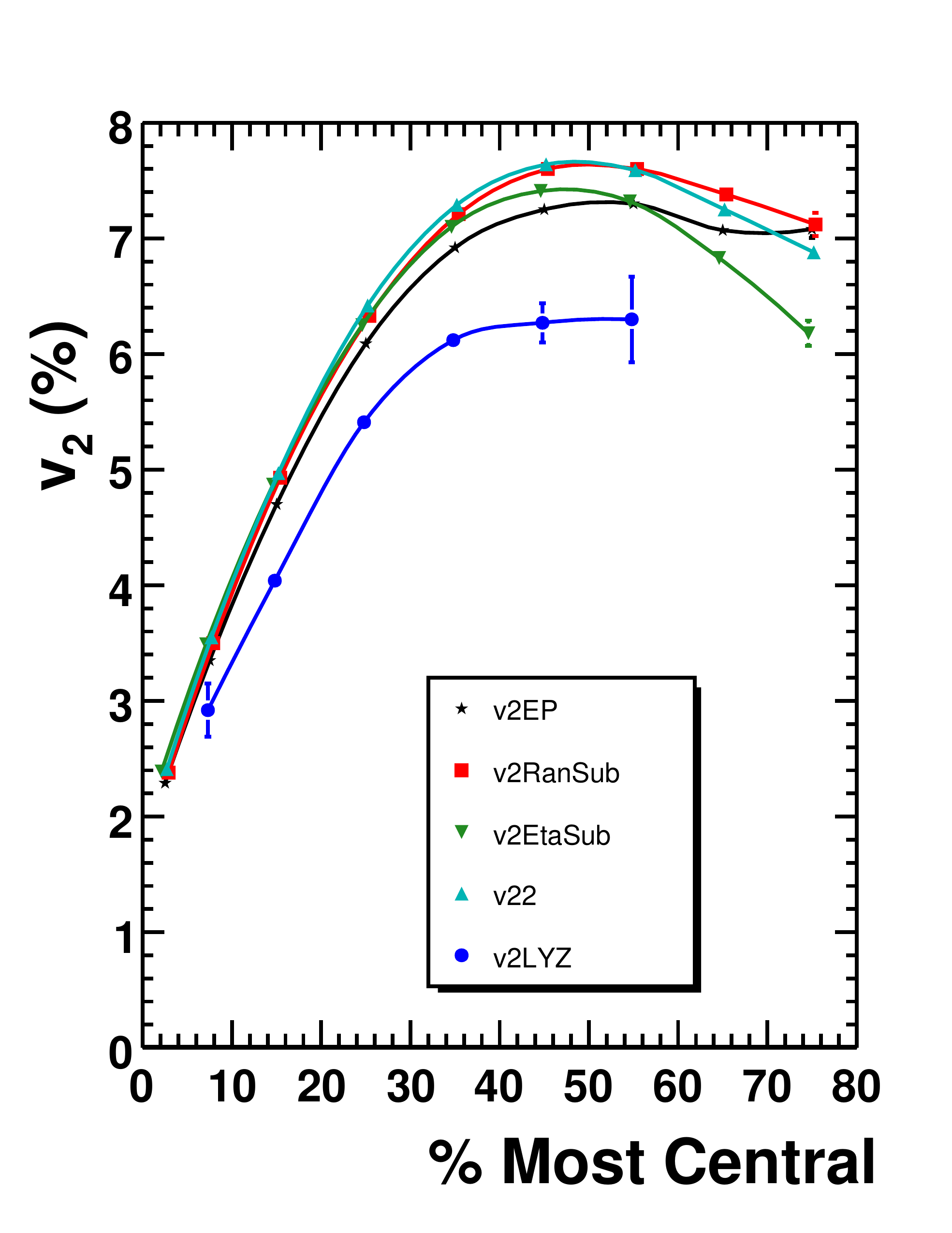}
 \caption{(Color online) The values of $v_2$ from various analysis methods vs centrality. Both the upper lines~\cite{Adams:2004bi} and the lower line~\cite{Abelev:2008ed} are STAR data. }
 \label{fig:pub_v}
\end{minipage}
\hspace{\fill}
\begin{minipage}[t]{0.48\textwidth}
\includegraphics[width=0.98\textwidth]{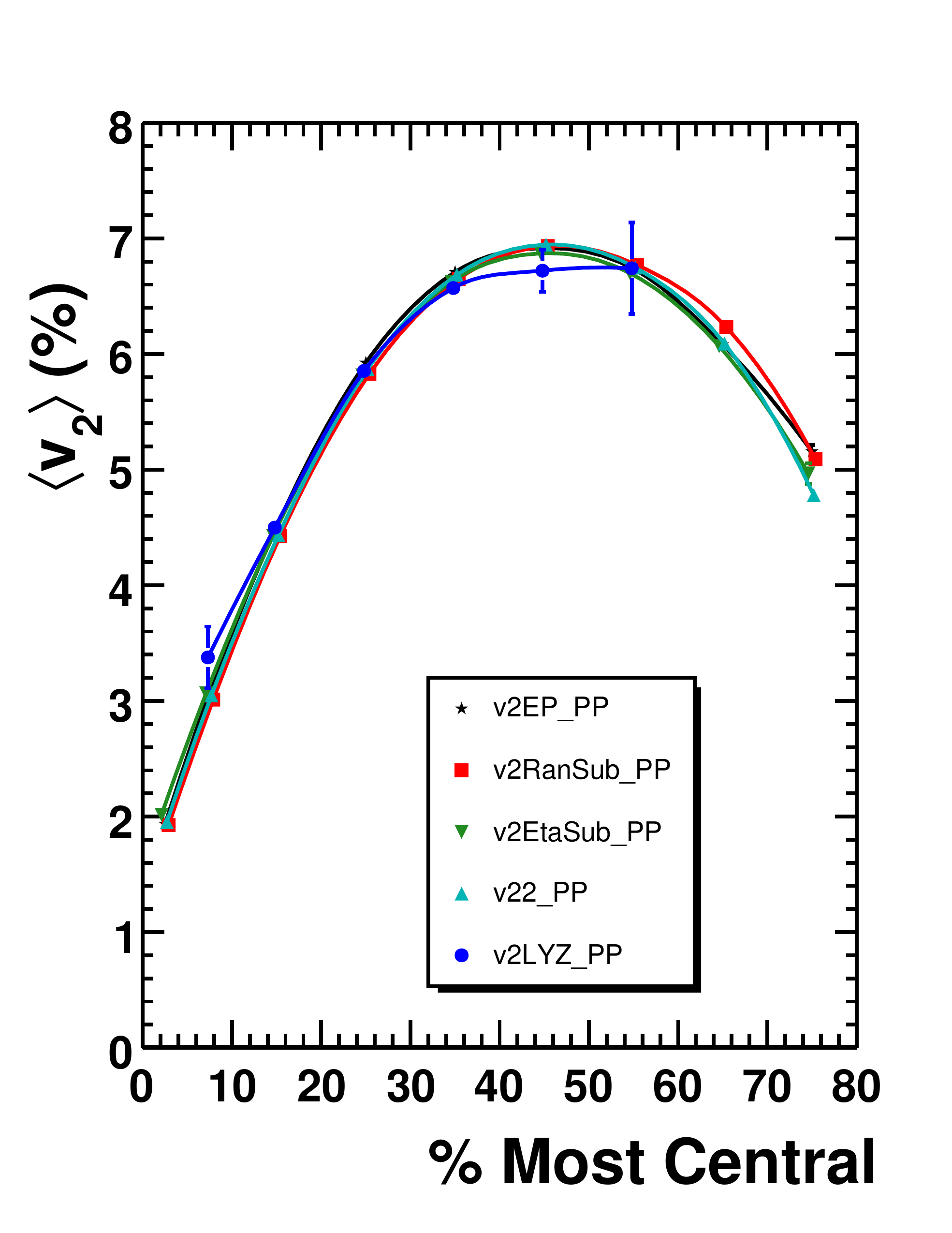}
 \caption{(Color online) The data from Fig.~\ref{fig:pub_v} corrected to $\mean{v_2}$ in the participant plane.}
  \label{fig:mean_v}
\end{minipage}
\end{figure*}
The published STAR data~\cite{Adams:2004bi,Abelev:2008ed} for the various methods are shown in Fig.~\ref{fig:pub_v}. The upper lines are from ``two-particle" correlation methods, and the lower line is from a multiparticle correlation method. The lower line values for $v_2\{{\rm LYZ}\}$ are thought to be in the reaction plane, if the fluctuations are Gaussian~\cite{Voloshin:2007pc}. The line for $v_2\{{\rm etaSub}\}$ is somewhat low for peripheral collisions because the gap in pseudorapidity reduces short-range nonflow correlations. Particularly puzzling is why the $\vtep$ line is lower than the other two-particle methods.

Correcting to $\mean{v}$ in the participant plane was done by using Eq.~(\ref{sumv2}) for $v_2\{2\}$, Eq.~(\ref{sumv4}) for $v_2\{{\rm LYZ}\}$, Eq.~(\ref{sumvep}) for \vtep, and Eq.~(\ref{sumvsub}) for $v_2\{{\rm ranSub}\}$ and $v_2\{{\rm etaSub}\}$. The results are shown in Fig.~\ref{fig:mean_v}. Since $v_2\{{\rm etaSub}\}$ is less affected by nonflow, the value of $\delta_{pp}$ used for it was multiplied by 0.5. In Fig.~\ref{fig:mean_v} the convergence of the two-particle, full event plane, and multiparticle results to one locus in the participant plane is remarkable. Even the shape of the $v_2\{{\rm etaSub}\}$ curve has changed to match the others with only one additional parameter. Previously we took the spread in the values in Fig.~\ref{fig:pub_v} as an estimate of the systematic uncertainty.

\subsection{CGC fluctuations}
\begin{figure*}[hbt]
\begin{minipage}[t]{0.48\textwidth}
\includegraphics[width=0.98\textwidth]{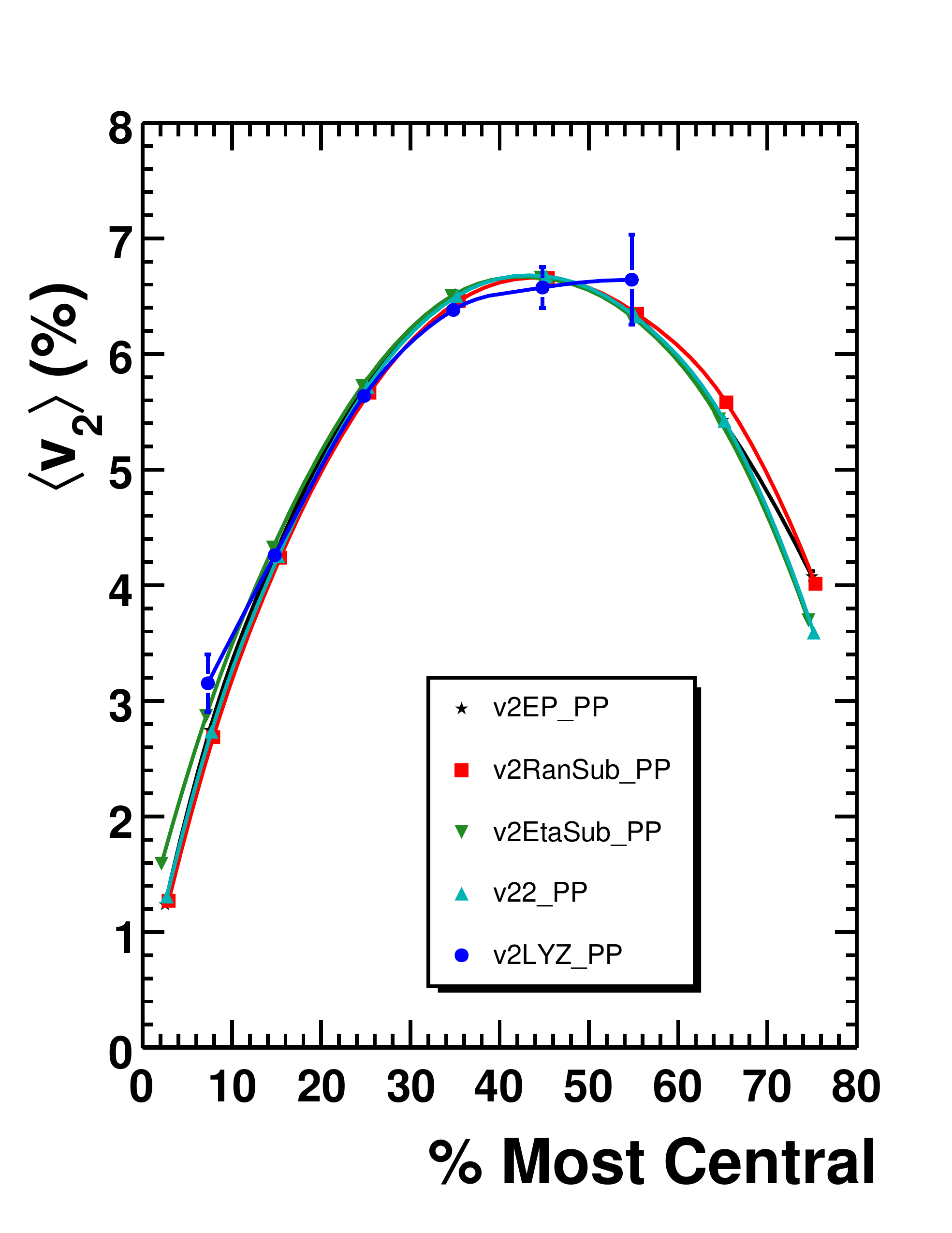}
 \caption{(Color online) The data from Fig.~\ref{fig:pub_v} corrected to $\mean{v_2}$ in the participant plane using CGC fluctuations and nonflow partly weighted with $N_{\rm bin}$. }
 \label{fig:mean_vCGC}
\end{minipage}
\hspace{\fill}
\begin{minipage}[t]{0.48\textwidth}
\includegraphics[width=0.98\textwidth]{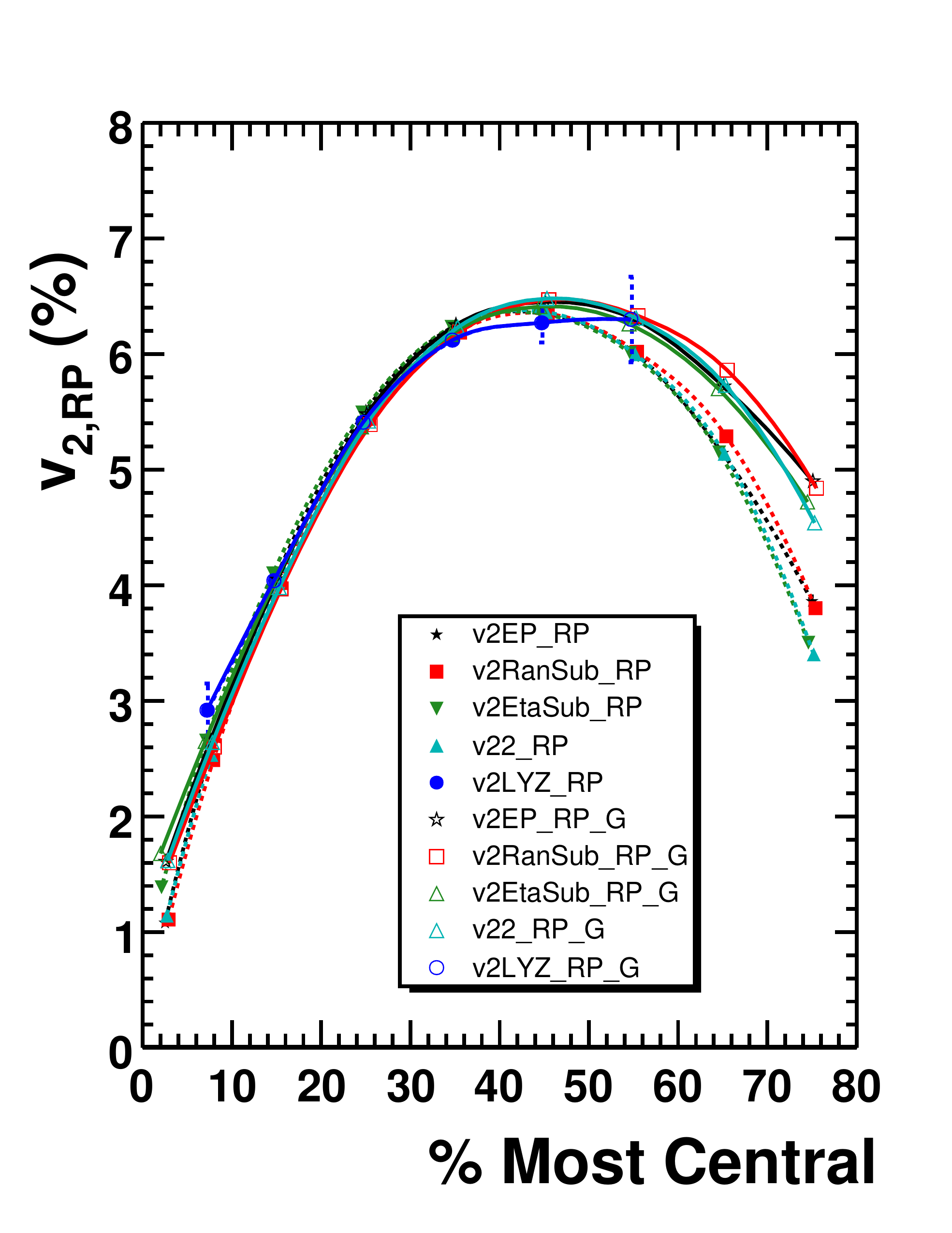}
 \caption{(Color online) The data from Figs.~\ref{fig:mean_v} and \ref{fig:mean_vCGC} corrected to $\mean{v_2}$ in the reaction plane. The solid lines are for the Glauber model of fluctuations and the dashed lines for the CGC model of fluctuations. }
 \label{fig:mean_vCGCRPboth}
\end{minipage}
\end{figure*}

To see how sensitive the convergence of the different methods is to
our assumptions for $\delta$ and $\sigma_v$, we also tried using
fluctuations in $\epspart$ from the Color Glass Condensate (CGC)
model~\cite{Drescher:2007ax}. In this model
$\sigma_{\eps}/\mean{\epspart}$ is roughly 30\% smaller, mainly
because $\mean{\epspart}$ is larger. Convergence of the
methods was not obtained because the values of $\sigma_{\rm tot}^2
\equiv \delta_2 + 2 \sigma_{v2}^2$ were too small. However, because
of hard scattering one might argue that for nonflow the number of
binary collisions is more important than the number of
participants. In fact, raising $\delta$ by weighting with the number
of binary collisions over the number of participants produced a
slight overcorrection. However, adjusting $\delta$ to be 70\%
weighted with $N_{\rm bin} / \npart$ and the rest just scaled with
$1/\npart$ brought $\sigtot$ down and produced reasonable
convergence: 
\begin{equation}
  \delta = \delta_{pp}\ 2\ [(x\ 2\ N_{\rm bin}/N_{\rm part}) + (1 - x)] /
	N_{\rm part}
\label{delCGC}
\end{equation}
with $x=0.7$. This assumes that for $p+p$ collisions, the number of participants is two and the number of binary interactions is one. Nonflow for $v_2\{{\rm etaSub}\}$ was reduced by multiplying by 0.7. With these assumptions the results are shown in Fig.~\ref{fig:mean_vCGC} in the participant plane. The convergence of the methods is also good.

\subsection{Reaction plane and parameters}
Using Eq.~(\ref{fluctv4}) and noting that $\mean{v} = \vpp$ and $v\{4\} \simeq \vrp$, the reaction plane values can be obtained from the participant plane values by~\cite{Voloshin:2007pc}
\begin{equation}
  \vpp^2 \simeq \vrp^2 + \sigma_{{\it v}}^2 \ ,
\label{RP}
\end{equation}
and, by using Eq.~(\ref{sigma_v}), 
\begin{equation}
  \vrp = \vpp\ \sqrt{1 - (\sigma_{\eps} / \mean{\eps})^2} \ .
\end{equation}
With this equation the values from Figs.~\ref{fig:mean_v} and \ref{fig:mean_vCGC} have been corrected to the reaction plane in 
Fig.~\ref{fig:mean_vCGCRPboth}. Thus, our two reasonable sets of assumptions about nonflow and fluctuations are not unique. At mid-centrality there is not much difference, but the graph illustrates the dependence on the systematic uncertainties in the assumptions that produce the corrections. $v_2$ in the reaction plane should go to zero at zero impact parameter. However, the first point in the graph is for 0--5\% centrality and there is some smearing in determining the centrality from the experimental multiplicity. Also, because of ground-state deformation of the Au nuclei, there could be some elliptic flow even at zero impact parameter~\cite{Filip:2007tj}. Since Eq.~(\ref{RP}) uses a Gaussian approximation for fluctuations in the participant plane, the \vrp\ values are not as reliable as the \vpp\ values, especially for peripheral centralities.

\begin{figure}[hbt]
\centerline{\includegraphics[width=.48\textwidth]{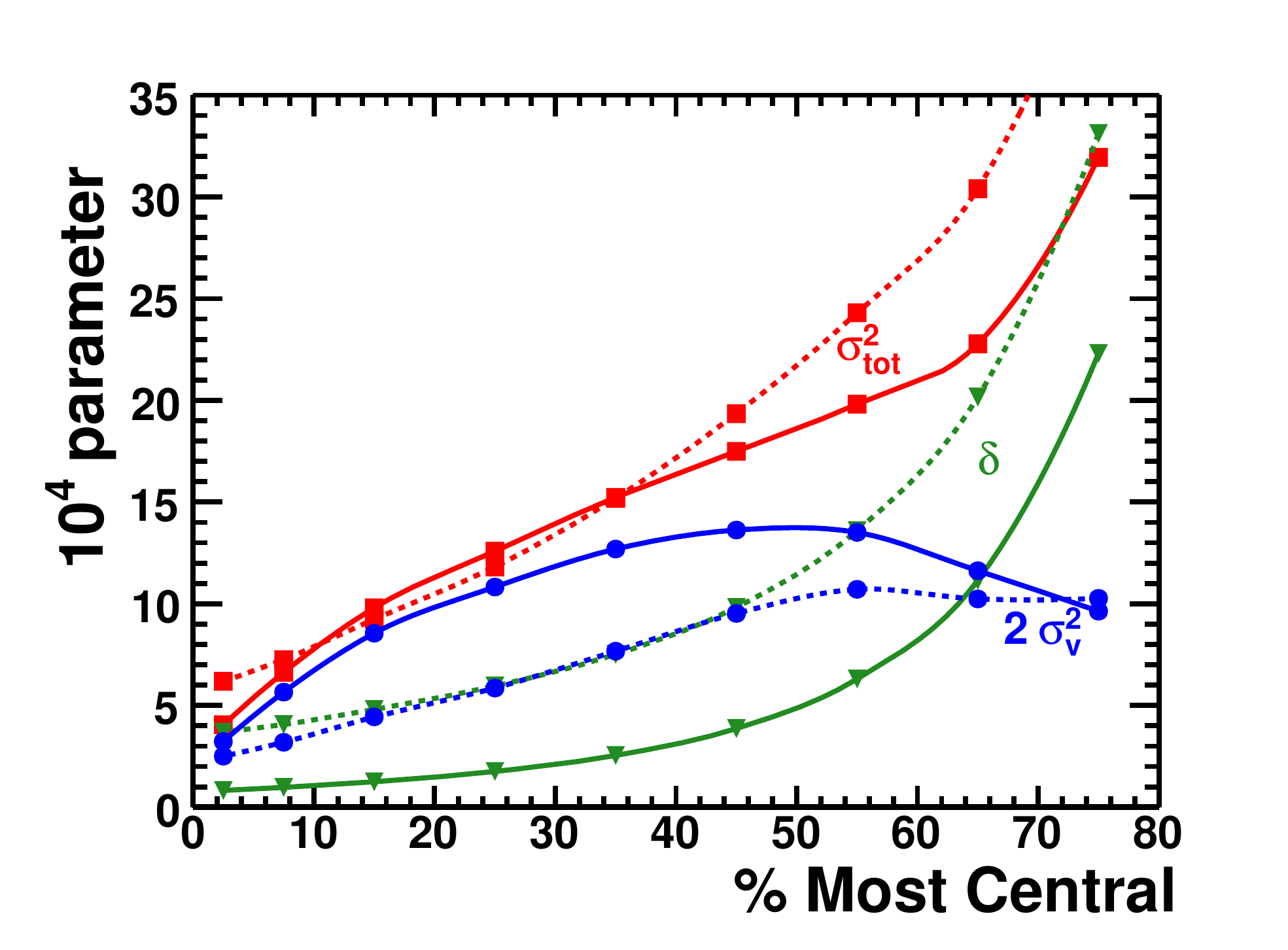}}
  \caption{(Color online) The nonflow and fluctuation parameters, derived from assumptions in the text, which were used to make the corrections of the various $v_2$ values to $\mean{v_2}$ in the participant plane and then to the reaction plane. $\sigma_{\rm tot}^2 \equiv \delta_2 + 2 \sigma_{v2}^2$.  See Table~\ref{tab:params}. The solid lines are for the Glauber model with $1/\npart$ scaling of $\delta$, and the dashed lines are for the CGC model with the addition of partial binary weighting for $\delta$. }
  \label{fig:parameters}
\end{figure}

The parameters used are shown in Fig.~\ref{fig:parameters} and in Table~\ref{tab:params}. The convergence depends on $\sigtot$ and is thus fixed from the experimental data by Eq.~(\ref{differences} top). It is about the same for the two sets of assumptions for centralities from 7.5\% to 50\%. However, the two assumptions differ in the proportion of $\sigma_{v2}$ and $\delta_2$ in $\sigtot$, and outside of this range the different assumptions give somewhat different results. For the unrealistic assumption of no nonflow, but only fluctuations, of course the Glauber and CGC models give very different results, and neither one shows convergence for the peripheral centralities.
\begin{table*}[hbt]
\centering
\begin{tabular}{rrrrrrrrr} \hline \hline
	& & & \ Glauber & & &  \ CGC \\
	bin \ & centrality  \ & mult \  & \vline \ $\sigma_{\eps}/\mean{\eps}$  \ &  \npart  \ & \ $\sigtot^2$ & \vline \  $\sigma_{\eps}/\mean{\eps}$  \ &  \nbin \ & \ $\sigtot^2$ \vline \\ \hline
     9 &  0 - 05\% & 961 & 55.5\% & 352 &  4.05 & 48.9\% & 1049 & 6.19 \\ 
     8 & 05 - 10\% & 819 & 50.2\% & 298 &  6.63 & 37.7\% &  825 & 7.25 \\ 
     7 & 10 - 20\% & 651 & 44.0\% & 232 &  9.80 & 31.7\% &  587 & 9.24 \\ 
     6 & 20 - 30\% & 468 & 38.2\% & 165 & 12.6  & 28.1\% &  364 & 11.8 \\ 
     5 & 30 - 40\% & 323 & 36.4\% & 114 & 15.2  & 28.3\% &  216 & 15.2 \\ 
     4 & 40 - 50\% & 214 & 36.0\% &  75 & 17.5  & 30.1\% &  120 & 19.3 \\ 
     3 & 50 - 60\% & 134 & 35.6\% &  46 & 19.8  & 31.7\% &   61 & 24.3 \\ 
     2 & 60 - 70\% & 76 & 34.1\% &  26 & 22.8  & 32.0\% &   28 & 30.4 \\ 
     1 & 70 - 80\% & 38 & 31.0\% &  13 & 31.9  & 32.0\% &   11 & 43.4 \\ \hline \hline
\end{tabular}
	\caption{For each centrality are shown the full event multiplicity~\cite{Adams:2004bi}, the standard deviation of Monte-Carlo Glauber $\epspart$ in percent of the mean~\cite{Hiroshi}, the number of participants~\cite{Adams:2004bi}, Glauber $\sigtot^2 \times 10^4$ as calculated here, the standard deviation of CGC $\epspart$ in percent of the mean~\cite{Drescher:2007ax}, the number of binary collisions~\cite{Adams:2004bi}, and CGC $\sigtot^2 \times 10^4$ as calculated here. The values of $\delta_2$ are given by Eq.~(\ref{delta}) for the Glauber model and Eq.~(\ref{delCGC}) for the CGC model. }
\label{tab:params}
\end{table*}

We also tried the extreme assumption of no fluctuations and calculated $\delta_2$ at each centrality from $v_2\{2\}^2 - v_2\{4\}^2$ as is indicated in Eq.~(\ref{differences} top) for $\sigma_v = 0$. The convergence of the methods for centralities from 7.5\% to 50\% was good since $v_2\{2\}$ and $v_2\{4\}$ are forced together. However, for peripheral collisions there was less convergence among the other values than shown in Fig.~\ref{fig:mean_vCGCRPboth}, and these values were lower than the other two sets of curves. As $\sigma_{v2}$ decreases from Glauber to CGC to zero, the more peripheral points decrease. That is because as $\sigma_{v2}$ decreases, $\delta_2$ must increase to compensate, and the more peripheral bins are most affected by $\delta_2$. Although we cannot rule out this no-fluctuation assumption, the convergence of the methods is not as good as for the other two cases.

%% file: summary.tex
\section{summary}
\label{sec:summary}

We have shown how the various experimental measures of elliptic flow are affected by fluctuations and nonflow, and we derived analytic equations which are leading order in $\sigma_v^2$ and $\delta$. For $\vsep$ and $\vep$ we have shown how the analytic values for fluctuations differ from simulations and a numerical integration of the distribution of $v$. We have transformed published data to the participant plane and then to the reaction plane using reasonable assumptions for fluctuations and nonflow. The convergence of the various experimental measurements is remarkable. We have shown this for two sets of assumptions, showing how the values depend on these assumptions. The convergence of the methods essentially fixes the value of $\sigtot$ from experimental data, but the separation into fluctuation and nonflow parts is not unique. To avoid both, better results for multiparticle correlations are needed. 

This procedure could also be applied to differential flow. Probably the relative fluctuations $\sigma_v/\mean{v}$, but not the nonflow, should be independent of pseudorapidity and transverse momentum. The nonflow as a function of \pt might be obtained from $p+p$ collisions as was done here for the integrated flow.

%% file: appendix.tex
\appendix
\section{Effects of fluctuations on \vep}
\label{sec:appendixA}

We derive the difference $\delta v$ between $\vep$ and $\mean{v}$ due to
fluctuations, to leading order in $\sigma_v^2$, assuming that nonflow
effects are negligible.  
Flow fluctuations modify both the numerator and the denominator of 
Eq.~(\ref{defvep}). 

\subsection{Small fluctuations}
\label{sec:appendixA1}

For small fluctuations, the relative change of
$\vep$ is obtained by taking the logarithm of Eq.~(\ref{defvep}) and 
differentiating 
\begin{equation}
\frac{\delta v}{\mean{v}}\equiv\frac{\vep-\mean{v}}{\mean{v}}=
\frac{\delta \mean{\cos (\phi-\Psi_R)}}{ \mean{\cos (\phi-\Psi_R)}}
-\frac{\delta R}{R}. 
\label{fluctnumdenom}
\end{equation}
The first term on the right-hand side is the contribution of
fluctuations to the correlation with the event plane; the second term
is the contribution of fluctuations to the resolution. We evaluate
these contributions in turn. 

The resolution parameter $\chi$ in Eq.~(\ref{resolution}) is
proportional to the flow $v$. If the analysis is done with unit
weights as in Eq.~(\ref{defq}), then $\chi=v\sqrt{N}$, where
$N$ is the number of particles in the event plane. More generally, we
write $\chi=r v$, where $r$ is a parameter depending on the details of
the analysis.  For a given value of $v$,
$\mean{\cos(\phi-\Psi_R)}=v{\cal  R}(rv)$. If $v$ fluctuates,
one must average this quantity   over the fluctuations of $v$. Using
Eq.~(\ref{averagef}) with  $f(v)=v{\cal R}(rv)$, the relative change
due to fluctuations is  
\begin{eqnarray}
\frac{\delta \mean{\cos (\phi-\Psi_R)}}{ \mean{\cos (\phi-\Psi_R)}}
&=& \frac{\sigma_v^2}{2}\frac{\frac{d^2}{dv^2}\left(v{\cal
 R}(rv\right))}{\mean{v}{\cal R}(r\mean{v})} \cr
&=& \frac{\sigma_v^2}{2\mean{v}^2}\frac{\chi\frac{d^2}{d\chi^2}\left(\chi{\cal
 R}(\chi\right))}{{\cal R}(\chi)}.
\label{defflnum}
\end{eqnarray}
In the second equality, $\chi\equiv r\mean{v}$ denotes the resolution parameter associated with $\mean{v}$. 
Using Eq.~(\ref{resolution}), one obtains after some algebra
\begin{eqnarray}
\frac{\delta \mean{\cos (\phi-\Psi_R)}}{ \mean{\cos (\phi-\Psi_R)}}
=\frac{\sigma_v^2}{2\mean{v}^2}
\left(1+\frac{I_0-I_1}{I_0+I_1}(1-2\chi^2)\right)
\label{flnum}
\end{eqnarray}
and $I_{0,1}$ is a shorthand notation for $I_{0,1}(\chi^2/2)$.

We now evaluate the second term in Eq.~(\ref{fluctnumdenom}), namely,
the shift in the resolution due to fluctuations.  
To estimate the resolution experimentally, one 
correlates two subevents $A$ and $B$. In the absence of fluctuations, 
$\mean{\cos(\Psi_A-\Psi_B)}={\cal R}(\chi_s)^2$, where 
$\chi_s$ is the resolution parameter of one subevent. 
With unit weights, $\chi_s=v\sqrt{N_s}$, where $N_s$ is
the number of particles in a subevent. More generally, one can write
$\chi_s=r_s v$, where $r_s$ is a parameter which depends on the
details of the analysis. The modification of the correlation due to fluctuations is evaluated
using Eq.~(\ref{averagef}), with  $f(v)={\cal R}(r_sv)^2$:
\begin{eqnarray}
\frac{\delta\mean{\cos(\Psi_A-\Psi_B)}}{\mean{\cos(\Psi_A-\Psi_B)}}
&=&
\frac{\sigma_v^2}{2}\frac{\frac{d^2}{dv^2}\left({\cal R}(r_s
v)^2\right)}{{\cal R}(r_s\mean{v})^2} \cr
&=& \frac{\sigma_v^2}{2\mean{v}^2}\frac{\chi_s^2\frac{d^2}{d\chi_s^2}\left({\cal
R}(\chi_s)^2\right)}{{\cal R}(\chi_s)^2}.
\label{resfluct1}
\end{eqnarray}
In the second equality, $\chi_s\equiv r_s\mean{v}$ denotes the
mean subevent resolution parameter. 

The resolution parameter of the subevent is determined experimentally 
by solving  Eq.~(\ref{defchis}). We denote by $\chi_s^{\rm exp}$ the 
solution of this equation and by $\delta\chi_s$ the shift due to
fluctuations: $\chi_s^{\rm
exp}=\chi_s+\delta \chi_s$. 
Differentiating Eq.~(\ref{defchis}), we obtain, to leading order in
$\delta\chi_s$,
\begin{equation}
\frac{{\cal R}'(\chi_s)\delta\chi_s}{{\cal R}(\chi_s)}=
\frac{1}{2}\frac{\delta\mean{\cos(\Psi_A-\Psi_B)}}
{\mean{\cos(\Psi_A-\Psi_B)}}
\label{denom0}
\end{equation}
Inserting Eq.~(\ref{resfluct1}), one obtains
\begin{equation}
\frac{\delta\chi_s}{\chi_s}=
\frac{\sigma_v^2}{2\mean{v}^2}\frac{\chi_s}{2{\cal R}(\chi_s){\cal R}'(\chi_s)}
\frac{d^2}{d\chi_s^2}\left({\cal
 R}(\chi_s)^2\right). 
\end{equation}
Using Eq.~(\ref{resolution}), one obtains after some algebra
\begin{equation}
\frac{\delta\chi_s}{\chi_s}=\frac{\sigma_v^2}{2\mean{v}^2}
\left(-2\chi_s^2+1+\frac{4 i_1^2}{i_0^2-i_1^2}\right) \ ,
\label{denom1}
\end{equation}
where $i_{0,1}$ is a shorthand notation for $I_{0,1}(\chi_s^2/2)$. 
The resolution parameter of the whole event, $\chi^{\rm exp}$, is
defined from the subevent resolution $\chi^{\rm exp}_s$ by 
\begin{equation}
\chi^{\rm exp}\equiv\sqrt{N/N_s}\chi^{\rm exp}_s=(r/r_s)\chi^{\rm
exp}_s
\end{equation}
Writing $\chi^{\rm exp}=\chi+\delta\chi$, where $\delta\chi$ is the
shift due to fluctuations, the resulting change $\delta R$ in the
resolution of the whole event is given by Eq.~(\ref{defresolution}): 
\begin{equation}
\frac{\delta R}{R}=\frac{\chi {\cal R}'(\chi)}{{\cal R}(\chi)}
\frac{\delta\chi}{\chi}=\frac{\chi {\cal R}'(\chi)}{{\cal R}(\chi)}
\frac{\delta\chi_s}{\chi_s}.
\label{denom2}
\end{equation}
Using Eq.~(\ref{resolution}), we obtain
\begin{equation}
\frac{\chi{\cal R}'(\chi)}{{\cal R}(\chi)}=
\frac{I_0-I_1}{I_0+I_1}.
\label{rprime}
\end{equation}
Inserting Eqs.~(\ref{denom1}) and (\ref{rprime}) into (\ref{denom2}),
we obtain
\begin{equation}
\frac{\delta R}{R}=
-\frac{I_0-I_1}{I_0+I_1}
\left(-2\chi_s^2+1+\frac{4
i_1^2}{i_0^2-i_1^2}\right)\frac{\sigma_v^2}{2\mean{v}^2}. 
\label{fldenom}
\end{equation}
Inserting Eqs.~(\ref{flnum}) and (\ref{fldenom}) into 
Eq.~(\ref{fluctnumdenom}), one obtains $\delta v/\mean{v}$. Finally,
using  
\begin{equation}
v\{{\rm EP}\}^2=(\mean{v}+\delta v)^2\simeq
\mean{v}^2+2\mean{v}^2\frac{\delta{v}}{\mean{v}}, 
\label{flvepbis}
\end{equation}
one obtains Eq.~(\ref{flvep}).

\subsection{Central collisions}
\label{sec:appendixA2}
Flow fluctuations are largest, in relative magnitude, for central
collisions. In this section, we derive exact formulas for the effect
of fluctuations on $\vep$ and $\vsep$ in central collisions.
By comparing these exact results with the approximate formulas derived 
above for small fluctuations, we will be able to assess the accuracy 
of the small-fluctuation approximations.

We assume that flow fluctuations result from Gaussian eccentricity 
fluctuations~\cite{Voloshin:2007pc}. The flow is given by $v=\sqrt{v_x^2+v_y^2}$, where the distribution of ${\bf v}=(v_x,v_y)$ is a
two-dimensional Gaussian
\begin{equation}
\label{gaussianv}
\frac{dn}{dv_x
dv_y}=\frac{1}{2\pi\sigma_0^2}\exp\left(-\frac{(v_x-v_0)^2+v_y^2}{2\sigma_0^2}\right). 
\end{equation}
Integrating over the azimuthal angle of ${\bf v}$, one recovers the
Bessel-Gaussian distribution Eq.~(\ref{BG}). 
From now on, we assume $v_0=0$, as expected by symmetry for
central collisions with no flow.
Then, the first two moments of the distribution are
$\langle v\rangle=\sqrt{\pi/2}\,\sigma_0$ and $\langle
v^2\rangle=2\sigma_0^2$. 
For a given value of ${\bf v}$, the distribution of the flow vector of
a subevent (A or B) is also a Gaussian centered around the direction
of ${\bf v}$~\cite{Voloshin:1994mz,Ollitrault:1993ba}:
\begin{equation}
p_{{\bf v}}({\bf q}_{A,B})=\frac{1}{\pi}\exp\left(-(
{\bf q}_{A,B}-{\bf v}\sqrt{N/2})^2\right). 
\end{equation}
The distribution of the flow vector of the whole event is given by a
similar equation, with $N$ instead of $N/2$. 
A factor $N$ comes from having $N$ particles in the 
event plane, and a factor $1/\sqrt{N}$ comes from the definition of the flow vector, Eq.~(\ref{defq}). The resolution ${\cal R}(\chi)$ in Eq.~(\ref{resolution}) is obtained by computing the average value of $\cos\Delta\Phi_R\equiv q_x/q$ with this distribution. When $v$ fluctuates, it is in fact easier to integrate first over $v$, then over $q$. One thus obtains the numerator of Eq.~(\ref{subs})
\begin{equation}
\mean{v{\cal R}(v\sqrt{N/2})}=\frac{\chi_s}{\sqrt{(\pi/4)+\chi_s^2}}\mean{v},
\label{b=0num}
\end{equation}
where $\chi_s\equiv\mean{v}\sqrt{N/2}$ is the average resolution
parameter of a subevent. The numerator of Eq.~(\ref{EP}) is given by a similar equation, with $\chi_s$ replaced by $\chi=\mean{v}\sqrt{N}$.

We now evaluate the correlation between subevents. 
Let ${\bf q}_A$ and ${\bf q}_B$ denote the flow vectors of subevents $A$
and $B$. Neglecting nonflow correlations, the joint probability
distribution of ${\bf q}_A$ and ${\bf q}_B$ is $p_{{\bf v}}(
{\bf q}_A)p_{{\bf v}}({\bf q}_B)$ for a given flow ${\bf v}$. Integrating over ${\bf v}$, one obtains the following probability distribution for $({\bf q}_A,{\bf q}_B)$:
\begin{eqnarray}
\label{distsubev}
\lefteqn{p({\bf q}_A,{\bf q}_B) = \frac{1}{\pi^2(1-C^2)}} \cr
&\times& \exp \left(-\frac{q_A^2 + q_B^2 - 
2C{\bf q}_A\cdot{\bf q}_B}{1-C^2} \right),
\end{eqnarray}
where 
\begin{equation}
C\equiv\frac{N\sigma_0^2}{1+N\sigma_0^2}
\label{defC}
\end{equation}
is the linear correlation between ${\bf q}_A$ and ${\bf q}_B$. 
The probability distribution Eq.~(\ref{distsubev}) is a correlated
Gaussian distribution, which is formally identical to the distribution
in the presence of nonflow effects~\cite{Borghini:2002mv,Lukasik:2006wx}. 
The relative angle between subevents, $\Delta\Phi$, is given by 
$\cos\Delta\Phi={\bf q}_A\cdot{\bf q}_B/(q_A q_B)$.
Integrating over ${\bf q}_A$ and ${\bf q}_B$, one obtains after some algebra
\begin{equation}
\langle [ \res (v\sqrt{N/2})]^2 \rangle=\mean{\cos\Delta\Phi}=
\frac{E(C^2)-(1-C^2)K(C^2)}{C}, 
\label{b=0den}
\end{equation}
where $K$ and $E$ are complete elliptic integrals of the first and
second kind. The value of $\mean{\cos\Delta\Phi}$ is approximately 
$(\pi/4)C$ for $C\ll 1$, and it is equal to 1 for $C=1$. 
The exact values of $\vsep$ and $\vep$ are obtained by
inserting Eqs.~(\ref{b=0num}) and (\ref{b=0den}) into
Eqs.~(\ref{subs}) and (\ref{EP}). The ratio of these exact results to
the approximate expressions of Eqs.~(\ref{flvsubep}) and (\ref{flvep}) is
plotted in Fig.~\ref{fig:b=0}.

\section{Effects of nonflow correlations on \vep}
\label{sec:appendixB}

The nonflow correlation is denoted by $\delta$ in
Eq.~(\ref{delDef}). In this Appendix, we denote it by $\delta_{\rm
nf}$ to avoid ambiguity, while $\delta X$ denotes 
the small change of an observable $X$ due to nonflow correlations. 
We derive the expression of $\vep$ to leading order in
$\delta_{\rm nf}$, neglecting flow fluctuations.

In the same way as fluctuations, nonflow effects contribute to both
the numerator and denominator of Eq.~(\ref{defvep}):
Eq.~(\ref{fluctnumdenom}) still holds, except that the shift is due to
nonflow instead of fluctuations. Nonflow effects give a
direct contribution to the correlation between the particle and the
event plane and to the correlation between subevents. In addition,
nonflow effects modify the distribution of the flow vector, which induces a change in the resolution parameter. We evaluate all these nonflow contributions separately when the flow vector is defined with unit weights, as in Eq.~(\ref{defq}). In practice, the analysis is often done with $\pt$ weights to increase the resolution. This case is more complex and will be discussed at the
end.

\subsection{Correction to the resolution parameter}

The normalized  probability distribution of the flow vector, defined
by Eq.~(\ref{defq}), is Gaussian:
\begin{equation}
p({{\bf q}})=\frac{1}{\pi\sigma^2}\exp\left(-\frac{(
{\bf q}-v\sqrt{N}{\bf e}_x)^2}{\sigma^2}\right), 
\label{centrallimit}
\end{equation}
where ${\bf e}_x$ is the unit vector along the true reaction plane, chosen as
the $x$-axis. 
In the absence of nonflow effects, $\sigma=1$ due to the normalization
factor $1/\sqrt{N}$ in Eq.~(\ref{defq}), and 
Eq.~(\ref{centrallimit}) reduces to 
\begin{eqnarray}
p({{\bf q}})&=&\frac{dN}{
qdqd\Psi_R}\cr
&=&\frac{1}{\pi}\exp\left(-q^2-\chi^2+2q\chi\cos\Psi_R\right).
\label{qdist}
\end{eqnarray}
Nonflow effects modify $\sigma$. Since the flow vector ${\bf q}$ in 
Eq.~(\ref{defq}) involves $N$ particles, the average value of $q^2$
involves $N^2$ correlated pairs. These pairs have nonflow
correlations, defined by Eq.~(\ref{delDef}). With the $1/\sqrt{N}$
normalization factor in Eq.~(\ref{defq}), one obtains

\begin{equation}
\sigma^2=1+N\delta_{\rm nf}.
\label{sigmanonflow}
\end{equation}
This change in $\sigma$ induces a change in the resolution
parameter~\cite{Borghini:2002mv}: 
\begin{equation}
\frac{\delta\chi}{\chi}=-\frac{\delta\sigma}{\sigma}=-\frac{N\delta_{\rm nf}}{2}.
\label{chinonflow}
\end{equation}
Similarly, the resolution parameter of subevents, $\chi_s$, is changed by
the amount
\begin{equation}
\frac{\delta\chi_{s}}{\chi_s}=-\frac{N_s\delta_{\rm nf}}{2}.
\label{chisnonflow}
\end{equation}

\subsection{Correlation with the event plane}
\label{s:nfnum}

Without nonflow effects, the correlation between the particle and the
event plane is 
\begin{equation}
\mean{\cos(\phi-\Psi_R)}= v {\cal R}(\chi)
\label{numeratorflow}
\end{equation}
Nonflow effects modify this equation in two different ways: 
1) the nonflow correlation between the particle and the event plane adds an extra term to the right-hand side;
2) $\chi$ is modified according to Eq.~(\ref{chinonflow}). 

We first evaluate the nonflow correlation between the particle and the
event plane. Let ${\bf u}\equiv (\cos\phi, \sin\phi)$ denote the unit vector of
the particle momentum. As shown in Ref.~\cite{Borghini:2002mv} in the case of momentum conservation, nonflow correlations between the flow vector and the
particle amount to shifting the flow vector by a small amount
proportional to ${\bf u}$. It can easily be shown that the shift is 
$\delta_{\rm nf}\sqrt{N}{\bf u}$, where a factor $N$ comes from having $N$
particles in the event plane and a factor $1/\sqrt{N}$ comes from the
definition of ${\bf q}$, Eq.~(\ref{defq}). The correlation between the particle and the event plane can be written as $\cos(\phi-\Psi_R)={\bf u}\cdot{\bf q}/q$. By shifting the flow vector and expanding to leading order in $\delta_{\rm nf}$, the resulting contribution to $\cos(\phi-\Psi_R)$ is  
\begin{eqnarray}
\delta\cos(\phi-\Psi_R)&=&\frac{\sqrt{N}\delta_{\rm
nf}}{q}\left(1-\frac{({\bf u}\cdot{\bf q})^2}{q^2}\right)\cr
&=&\frac{\sqrt{N}\delta_{\rm nf}}{q}\sin^2(\phi-\Psi_R). 
\end{eqnarray}
Averaging over events, $\sin^2(\phi-\Psi_R)$ gives $1/2$. 
The average value of $1/q$ is computed using Eq.~(\ref{qdist}):
\begin{equation}
\left\langle \frac{1}{q}\right\rangle= \int \frac{1}{q} \frac{dN}{q dq
 d\Psi_R} q dq d\Psi_R=
\frac{\sqrt{\pi}}{
2}e^{-\chi^2/2}2  I_0.
\end{equation}
To obtain the relative change due to nonflow effects, we
divide by Eq.~(\ref{numeratorflow}), where ${\cal R}(\chi)$ is given
by Eq.~(\ref{resolution}), and $\chi=v\sqrt{N}$:
\begin{equation}
\frac{\delta\mean{\cos(\phi-\Psi_R)}}{\mean{\cos(\phi-\Psi_R)}}
=
\frac{2I_0}{(I_0+I_1)}\frac{\delta_{\rm nf}}{2 v^2}
=
\left(1+\frac{I_0-I_1}{I_0+I_1}\right)\frac{\delta_{\rm nf}}{2 v^2}
\label{nfnumerator1}
\end{equation}
We now evaluate the second contribution, arising from the modification
of $\chi$, Eq.~(\ref{chinonflow}). The resulting change is
\begin{equation}
\frac{\delta\mean{\cos(\phi-\Psi_R)}}{\mean{\cos(\phi-\Psi_R)}}
=\frac{\delta{\cal R}}{\cal R}
=\frac{{\cal R}'(\chi)}{{\cal R}(\chi)}\delta\chi 
=-\frac{\chi{\cal R}'(\chi)}{{\cal R}(\chi)}\frac{N\delta_{\rm
   nf}}{2}.
\end{equation}
Using Eq.~(\ref{rprime}) and $N=\chi^2/v^2$, this becomes
\begin{equation}
\frac{\delta\mean{\cos(\phi-\Psi_R)}}{\mean{\cos(\phi-\Psi_R)}}
=-\frac{I_0-I_1}{I_0+I_1} \chi^2
\frac{\delta_{\rm nf}}{2 v^2} \ .
\label{nfnumerator2}
\end{equation}
Adding the contributions from Eqs.~(\ref{nfnumerator1}) and
(\ref{nfnumerator2}), we obtain the total nonflow contribution to the 
correlation between the particle and the event plane:
\begin{equation}
\frac{\delta\mean{\cos(\phi-\Psi_R)}}{\mean{\cos(\phi-\Psi_R)}}
=
\left(1+\frac{(I_0-I_1)}{(I_0+I_1)}(1-\chi^2)\right)\frac{\delta_{\rm
nf}}{2 v^2} \ .
\label{nfnumerator}
\end{equation}

\subsection{Resolution correction}

We now derive the modification of the resolution correction due to
nonflow effects. As in Sec.~\ref{s:nfnum}, there are two nonflow contributions: the first contribution is the nonflow correlation between the subevents;
the second modification arises from the modification of the width of
the flow vector distribution, Eq.~(\ref{chisnonflow}). 

We first derive the correlation between subevents due to
nonflow effects. Let ${\bf q}_A$ and ${\bf q}_B$ denote the flow vectors of subevents $A$ and $B$. The joint probability distribution of ${\bf q}_A$ and
${\bf q}_B$ is~\cite{Borghini:2002mv}
\begin{widetext}
\begin{equation}
\frac{dN}{d^2{\bf q}_A d^2{\bf q}_B}=p({\bf q}_A)p({\bf q}_B)
\left(1+2N_s\delta_{\rm nf} ({\bf q}_A-\chi{\bf e}_x)\cdot ({\bf
q}_B-\chi{\bf e}_x)\right),
\end{equation}
where $p({\bf q}_A)$ is defined by Eq.~(\ref{qdist}) (except that
$\chi$ is replaced with $\chi_s$), and the term proportional to 
$\delta_{\rm nf}$ is the nonflow correlation between subevents. The factor $N_s$
is due to the correlation being $N_s$ times stronger between
subevents than between individual particles. One then computes $\langle\cos(\Psi_A-\Psi_B)\rangle$ with this probability distribution. The nonflow contribution reads
\begin{equation}
\delta
\mean{\cos(\Psi_A-\Psi_B)}=2N_s\delta_{\rm nf}\left(
\mean{(q\cos \Psi_A-\chi)\cos \Psi_A}^2+
\mean{q\sin^2 \Psi_A}^2\right),
\end{equation}
\end{widetext}
where angular brackets on the right-hand side denote average values,
which are taken with the probability distribution
Eq.~(\ref{qdist}). These averages are easily evaluated as 
\begin{equation}
\mean{(q\cos \Psi_A-\chi)\cos \Psi_A}=
\frac{\sqrt{\pi}}{2}e^{-\chi_s^2/2}\frac{1}{2}\left(
i_0 -i_1 
\right)
\end{equation}
and 
\begin{equation}
\mean{q\sin^2 \Psi_A}=
\frac{\sqrt{\pi}}{2}e^{-\chi_s^2/2}\frac{1}{2}\left(
i_0 +i_1 
\right).
\end{equation}
One thus obtains 
\begin{equation}
\delta\mean{\cos(\Psi_A-\Psi_B)}
=N_s\delta_{\rm nf}\left(\frac{\sqrt{\pi}}{2}e^{-\chi_s^2/2}\right)^2
\left(i_0^2 +i_1^2 \right).
\label{nonflowsubevents}
\end{equation}
In the absence of nonflow effects,
$\mean{\cos(\Psi_A-\Psi_B)}={\cal R}(\chi_s)^2$, where ${\cal
R}(\chi_s)$ is given by Eq.~(\ref{resolution}). This gives the
relative variation
\begin{equation}
\frac{\delta\mean{\cos(\Psi_A-\Psi_B)}}{\mean{\cos(\Psi_A-\Psi_B)}}=
\frac{i_0^2+i_1^2}{(i_0+i_1)^2}\frac{\delta_{\rm nf}}{2 v^2},
\end{equation}
where we have used $N_s/\chi_s^2=1/v^2$. Nonflow effects introduce a bias $\delta\chi_s$ in the estimate of $\chi_s$, the resolution parameter of the subevent. This bias is given by Eq.~(\ref{denom0}): 
\begin{equation}
\frac{\delta\chi_s}{\chi_s}=
\frac{\delta\mean{\cos(\Psi_A-\Psi_B)}}{\mean{\cos(\Psi_A-\Psi_B)}}
\frac{{\cal R}(\chi_s)}{\chi_s {\cal R}'(\chi_s)}=
\frac{(i_0^2+i_1^2)}{(i_0^2-i_1^2)}
\frac{\delta_{\rm nf}}{2 v^2},
\label{deltachisnf1}
\end{equation}
where we have used Eq.~(\ref{rprime}), with $\chi_s$ instead of
$\chi$. The second effect is the modification of $\chi_s$ from the increase 
of the width of the distribution of the flow vector,
Eq.~(\ref{chisnonflow}). Writing $N_s=\chi_s^2/v^2$ and adding this
contribution to Eq.~(\ref{deltachisnf1}), we obtain
\begin{equation}
\frac{\delta\chi_s}{\chi_s}=
\left(\frac{i_0^2+i_1^2}{i_0^2-i_1^2}-\chi_s^2\right)
\frac{\delta_{\rm nf}}{2 v^2}
=
\left(1-\chi_s^2+\frac{2i_1^2}{i_0^2-i_1^2}\right)
\frac{\delta_{\rm nf}}{2 v^2} .
\label{deltachisnf}
\end{equation}
The relative correction to the resolution is then given by 
Eqs.~(\ref{denom2}) and (\ref{rprime}):
Inserting Eq.~(\ref{deltachisnf}),we obtain
\begin{equation}
\frac{\delta R}{R}
=\frac{I_0-I_1}{I_0+I_1}
\left(1-\chi_s^2+\frac{2i_1^2}{i_0^2-i_1^2}\right)
\frac{\delta_{\rm nf}}{2 v^2}. 
\label{nfdenominator}
\end{equation}

The relative change $\delta v/v$ is obtained from 
Eqs.~(\ref{fluctnumdenom}), using the results from
Eqs.~(\ref{nfnumerator}) and (\ref{nfdenominator}). 
Equation.~(\ref{nonflowep}) is finally obtained by using 
Eq.~(\ref{flvepbis}). 

\subsection{Weights}

We finally discuss the case where the flow analysis is done with
weights. This means that Eq.~(\ref{defq}) is replaced with 
\begin{eqnarray}
  {\bf q}\cos\Psi_R =\frac{\bf Q}{\sqrt{N}}\cos\Psi_R&=&\frac{1}{\sqrt{N}}\sum_{j=1}^N w_j\cos\phi_j \cr
  {\bf q}\sin\Psi_R=\frac{\bf Q}{\sqrt{N}}\sin\Psi_R &=&\frac{1}{\sqrt{N}}\sum_{j=1}^N w_j\sin\phi_j,
\label{defqw}
\end{eqnarray}
where $w_j$ is a weight which may depend on transverse momentum,
rapidity, and mass. Using appropriate weights increases the
resolution. The optimal weight is $w_j\propto
v_2$~\cite{Borghini:2001vi}. A standard choice for elliptic flow at
RHIC is $w=\pt$ up to 2 $\GeVc$ and flat above that.  

Our discussion of fluctuations in Appendix~\ref{sec:appendixA} is
independent of which weights are used. For nonflow effects, weights
matter. The problem is that the various nonflow terms listed in 
this Appendix are not all weighted in the same way. 
More specifically, the correlation between subevents will get weights
from particles from both subevents, while the correlation between the
particle and the event plane only gets one weight. 
We denote $\delta_{\rm full}$ as the nonflow correlation with one weight and
$\delta_{\rm sub}$ as the nonflow correlation with two weights. One must replace $\delta_{\rm nf}$ with $\delta_{\rm full}$ in
Eq.~(\ref{nfnumerator1}) and with $\delta_{\rm sub}$ in
Eqs.~(\ref{nfnumerator2}) and (\ref{nfdenominator}). 
Equation~(\ref{nonflowep}) is then replaced by
\begin{eqnarray}
\lefteqn{\vep^2 = \mean{v}^2
+ \left(1-\frac{I_0-I_1}{I_0+I_1}\right)\delta_{\rm full}} \cr
&-& \frac{I_0-I_1}{I_0+I_1}\left(\chi^2-\chi_s^2+\frac{2
 i_1^2}{(i_0^2-i_1^2)}\right) \delta_{\rm sub}.
 \label{nonflowepw}
\end{eqnarray}